# QUANTUM PHYSICS EXPLORING GRAVITY IN THE OUTER SOLAR SYSTEM: THE SAGAS PROJECT

P. Wolf[1], Ch. J. Bordé[1], A. Clairon[1], L. Duchayne[1], A. Landragin[1], P. Lemonde[1], G. Santarelli[1], W. Ertmer[2], E. Rasel[2], F.S. Cataliotti[3], M. Inguscio[3], G.M. Tino[3], P. Gill[4], H. Klein[4], S. Reynaud[5], C. Salomon[5], E. Peik[6], O. Bertolami[7], P. Gil[7], J. Páramos[7], C. Jentsch[8], U. Johann[8], A. Rathke[8], P. Bouyer[9], L. Cacciapuoti[10], D. Izzo[10], P. De Natale[11], B. Christophe[12], P. Touboul[12], S.G. Turyshev[13], J.D. Anderson[14], M.E. Tobar[15], F. Schmidt-Kaler[16], J. Vigué[17], A. Madej[18], L. Marmet[18], M-C. Angonin[1], P. Delva[1], P. Tourrenc[19], G. Metris[20], H. Müller[21], R. Walsworth[22], Z.H. Lu[23], L.J. Wang[23], K. Bongs[24], A. Toncelli[25], M. Tonelli[25], H. Dittus[26], C. Lämmerzahl[26], P. Laporta[27], J. Laskar[28], A. Fienga[29], F. Roques[30], K. Sengstock[31], G. Galzerano[32]

[1] *SYRTE, Observatoire de Paris, CNRS, France*
[2] *Leibniz Universität Hannover, Germany*
[3] *LENS, Universita' di Firenze, INFN, Italy*
[4] *National Physical Laboratory, United Kingdom*
[5] *Laboratoire Kastler Brossel, CNRS, France*
[6] *Physikalisch-Technische Bundesanstalt, Germany*
[7] *Instituto Superior Técnico, Portugal*
[8] *EADS-Astrium, Friedrichshafen, Germany*
[9] *Laboratoire Charles Fabry de l'Institut d'Optique, CNRS, France*
[10] *ESA, ESTEC, The Netherlands*
[11] *INOA-CNR, Firenze, Italy*
[12] *ONERA, Chatillon, France*
[13] *NASA, Jet Propulsion Laboratory, USA*
[14] *Jet Propulsion Laboratory, California Institute of Technology, USA*
[15] *University of Western Australia, Australia*
[16] *Universität Ulm, Germany*
[17] *LCAR, Université Toulouse III, CNRS, France*
[18] *National Research Council of Canada, CNRC, Canada*
[19] *ERGA/LERMA, Université Paris 6, France*
[20] *Géosciences Azur, Observatoire de la Côte d'Azur, CNRS, France*
[21] *Stanford University, USA*
[22] *Harvard University, USA*
[23] *Universität Erlangen, Germany*
[24] *University of Birmingham, United Kingdom*
[25] *NEST, INFM-CNR, Universita' di Pisa, Italy*
[26] *ZARM, University of Bremen, Germany*
[27] *Politecnico di Milano, Italy*
[28] *IMCCE, Observatoire de Paris, CNRS, France*
[29] *Observatoire de Besançon, France*
[30] *LESIA, Observatoire de Paris, CNRS, France*
[31] *Universität Hamburg, Germany*
[32] *CNR, IFN-Sez. Milano, Italy*

**Contact:** peter.wolf@obspm.fr

*Abstract:* We summarise the scientific and technological aspects of the SAGAS (Search for Anomalous Gravitation using Atomic Sensors) project, submitted to ESA in June 2007 in response to the Cosmic Vision 2015-2025 call for proposals. The proposed mission aims at flying highly sensitive atomic sensors (optical clock, cold atom accelerometer, optical link) on a Solar System escape trajectory in the 2020 to 2030 time-frame. SAGAS has numerous science objectives in fundamental physics and Solar System science, for example numerous tests of general relativity and the exploration of the Kuiper belt. The combination of highly sensitive atomic sensors and of the laser link well adapted for large distances will allow measurements with unprecedented accuracy and on scales never reached before. We present the proposed mission in some detail, with particular emphasis on the science goals and associated measurements.

## 1. INTRODUCTION

The SAGAS mission will study all aspects of large scale gravitational phenomena in the Solar System using quantum technology, with science objectives in fundamental physics and Solar System exploration. It will contribute to the search for answers to some of the major questions of relevance to present day physics and space science:

- SAGAS will carry out a large number of tests of fundamental physics, and gravitation in particular, at scales only attainable in a deep space experiment (sect. 2.2., 2.3., 2.4., 2.7., 2.8.). Given the scale and sensitivity of the SAGAS measurements, this will deeply probe the known laws of physics, with the potential for a major discovery in an area where many modern unification theories hint towards new physics. The unique combination of onboard instruments (atomic clock, atomic absolute accelerometer, optical link to ground) will allow for a two to five orders of magnitude improvement on many tests of special and general relativity, as well as a detailed exploration of a possible anomalous scale dependence of gravitation.
- SAGAS will provide detailed information on the Kuiper belt mass and mass distribution, the largely unexplored remnant of the circumsolar disk where the giant planets of the Solar System formed (see sect. 2.5. and 2.5.1). Additionally it will allow the precise determination of the mass of one or several Kuiper belt objects and possibly discover new ones (sect. 2.5.2.).
- SAGAS will carry out a measurement of the mass and mass distribution of the Jupiter system with unprecedented accuracy (sect. 2.6.).

This large spectrum of objectives makes SAGAS a unique combination of exploration and science, with a strong basis in both programs. The involved large distances (up to 53 AU) and corresponding large variations of gravitational potential combined with the high sensitivity of SAGAS instruments serve both purposes equally well. For this reason, SAGAS brings together traditionally distant scientific communities ranging from atomic physics through experimental gravitation to planetology and Solar System science.

The payload will include an optical atomic clock optimised for long term performance, an absolute accelerometer based on atom interferometry and a laser link for ranging, frequency comparison and communication. The complementary instruments will allow highly sensitive measurements of all aspects of gravitation via the different effects of gravity on clocks, light, and the free fall of test bodies, thus effectively providing effectively a detailed gravitational map of the outer Solar System whilst testing all aspects of gravitation theory to unprecedented levels. The detail on the different science objectives can be found in the appropriate sub-sections of sect. 2.

The SAGAS accelerometer is based on cold Cs atom technology derived to a large extent from the PHARAO space clock built for the ACES mission [30]. The PHARAO engineering model has recently been tested with success, demonstrating the expected performance and robustness of the technology. The accelerometer will only require parts of PHARAO (cooling and trapping region) thereby significantly reducing mass and power requirements. The expected sensitivity of the accelerometer (cf Tab. 2-1) is $1.3 \times 10^{-9}$ m/s$^2$ Hz$^{-1/2}$ with an absolute accuracy (bias determination) of $5 \times 10^{-12}$ m/s$^2$, the latter being crucial for many of the science objectives.

The SAGAS clock will be an optical clock based on trapped and laser cooled single ion technology as pioneered in numerous laboratories around the world. In the present baseline it will be based on a Sr$^+$ ion with a clock wavelength of 674 nm. The assumed stability of the SAGAS clock is $1 \times 10^{-14}/\sqrt{\tau}$ (with $\tau$ the integration time), with an accuracy in realising the unperturbed ion frequency of $1 \times 10^{-17}$. The best optical single ion ground clocks presently show stabilities slightly better ($3 \times 10^{-15}/\sqrt{\tau}$) than the one assumed for the SAGAS clock, and only slightly worse accuracies ($2 \times 10^{-17}$). So the technology challenges facing SAGAS are not so much the required performance, but the development of reliable and space qualified systems, with reduced mass and power consumption.

The optical link is using a high power (1 W) laser locked to the narrow and stable frequency provided by the optical clock, with coherent heterodyne detection on the ground and on board the S/C. It serves the multiple purposes of comparing the SAGAS clock to ground clocks, providing highly sensitive Doppler measurements for navigation and science, and allowing data transmission together with timing and coarse ranging. It is based on a 40 cm space telescope and 1.5 m ground telescopes (similar to lunar laser ranging stations). The main challenges of the link will be the required pointing accuracy (0.3") and the availability of space qualified, robust 1 W laser sources at 674 nm. Quite generally, laser availability and reliability will be the key to achieving the required technological performances, for the clock as well as the optical link (see sect. 6.2.).

For this reason a number of different options have been considered for the clock/link laser wavelength (see sect. 3.2.), with several other ions that could be equally good candidates (*e.g.* Yb$^+$ @ 435 nm and Ca$^+$ @ 729 nm). Given present laser technology, Sr$^+$ was preferred, but this choice could be revised depending on laser developments over the next years. We also acknowledge the possibility that femtosecond laser combs might be developed for space applications in the near future, which would open up the option of using either ion with existing space qualified 1064 nm Nd:YAG lasers for the link.

More generally, SAGAS technology takes advantage of important heritage from cold atom technology used in PHARAO and laser link technology designed for LISA (Lasers Ineterferometric Space Antenna). It will provide an excellent opportunity to develop those technologies for general use, including development of the ground segment (Deep Space Network telescopes and optical clocks), that will allow such technologies to be used in many other mission configurations for precise timing, navigation and broadband data transfer throughout the Solar System.

In summary, SAGAS offers a unique opportunity for a high profile deep space mission with a large spectrum of science objectives in Solar System exploration and fundamental physics, and the potential for a major breakthrough in our present conception of physics, the Solar System and the universe as a whole.

## 2. SCIENCE OBJECTIVES

SAGAS offers the possibility to achieve high priority Fundamental Physics and Solar System objectives, thus combining science and exploration in a unique way. One would expect that the corresponding large number of equally important objectives will require some compromises on trajectory design and data acquisition, but it is surprising to note that almost no trade-offs are required, as the different requirements are largely compatible.

Trajectory design plays an important role in many of the Solar System objectives, but almost no role on most fundamental physics goals, for which a large variation of gravitational field and large distance from the Sun is required, with no particular preferred direction (apart from a weak constraint on the general flight direction from the Lorentz Invariance (LI) test, and the requirement for at least one occultation for the PPN test). This leaves the possibility to optimise the trajectory for the Solar System objectives without significantly affecting the fundamental physics results.

Data analysis will then be tailored to share the time between Solar System science and fundamental physics. For example during close planetary or Kuiper Belt Object (KBO) fly-bys, the measurements will yield information on the planet or KBO whilst leaving ample time during the rest of the mission (when far from known objects) for the fundamental physics objectives. Designing an optimal trajectory will be one of the tasks of more detailed studies, but a trajectory satisfying all objectives can certainly be found, given the large space of possible options.

In the following sections we investigate the different science objectives based on "raw" measurement uncertainties of the different observables (section 2.1.). This said, we acknowledge that the actual data analysis will consist of fitting the different models under investigation together with models for known perturbations to the measurements, thus obtaining more sensitive results on all model parameters. Corresponding simulations could yield more reliable estimates of SAGAS performance concerning the different objectives and some information on correlations between them. Such simulations are beyond the scope of this overview, but we note that they should in most cases lead to more favourable estimates on the achievable results than the rough estimates provided here.

### 2.1. Measurements and Observables

SAGAS will provide three fundamental measurements: the accelerometer readout and the two frequency differences (measured on ground and on board the S/C) between the incoming laser signal and the local optical clock. Auxiliary measurements are the timing of arriving signals on board and on the ground that are used for ranging and time tagging of data. The high precision science observables will be deduced from the fundamental measurements by combining the measurements to obtain information on either the frequency difference between the clocks or the Doppler shift of the transmitted signals (see section 3.3.). The latter gives access to the relative satellite-ground velocity, from which the gravitational trajectory of the satellite can be deduced by correcting non-gravitational accelerations ($\mathbf{a}_{NG}$) using the accelerometer readings. Then the three science observables are

*Relative frequency:* $\quad y \equiv \partial_t \tau_S - \partial_t \tau_G$
*Doppler shift:* $\quad D_\nu \equiv (\nu_r - \nu_e)/\nu_0$ $\qquad$ (2-1)
*Non-gravitational acceleration:* $\quad \mathbf{a}_{NG}$

where $\tau_i$ is proper time at the position of the space and ground clock respectively, $t$ is coordinate time, $\nu_i$ is the received, emitted, and nominal proper frequency of a photon, and $D_\nu$ is corrected for non-gravitational satellite motion. From the observables we obtain the quantities of interest that give access to the science objectives: S/C gravitational motion, S/C proper time evolution, and light propagation.

In the following, we assume that Earth station motion and its local gravitational potential can be known and corrected to uncertainty levels below $10^{-17}$ in relative frequency (<10 cm on geocentric distance), which can certainly be achieved for the time varying parts of the potential and is only a factor 3 less than the best reported absolute determination [1] with still some room for improvement (*e.g.* the potential on the geoid is presently known to better than 5 cm [2], with improvements to about 1 cm expected from the GOCE mission [39]). For the Solar System parameters this requires $10^{-9}$ relative uncertainty for the ground clock parameters (*GM* and *r* of Earth), also achieved at present [2], and less stringent requirements for the S/C. A particular situation, however, is the close approach of planets, treated in detail in sect. 2.6.

The raw frequency measurements (on board and on the ground) can be combined in two fundamental ways (see section 3.3.). Their sum yields sensitivity to the relative S/C – Earth velocity via the first order Doppler effect with suppressed sensitivity to the clock noise. Their difference yields sensitivity to the clock frequency difference with suppressed sensitivity to the motion of the S/C and ground station. As a consequence, gravitational trajectory restitution will be determined by tropospheric and clock noise at high frequencies and S/C accelerometer noise at low frequencies ($< 10^{-4}$ Hz). Note that noise in the determination of the ground station motion plays a non-negligible role at intermediate frequencies, but is below the S/C acceleration noise at low frequencies, when using modern positioning techniques (Global Satellite Navigation Systems, Satellite Laser Ranging, Very Long Baseline Interferometry). For long term integration and the determination of an acceleration bias, the limiting factor will then be the accelerometer noise and absolute uncertainty (bias determination), also shown in Tab. 2-1. More generally, modelling of non-gravitational accelerations will certainly allow some improvement on the long term limits imposed by the accelerometer noise and absolute uncertainty, but is not taken into account in Tab. 2-1.

| | **Noise PSD / Hz$^{-1}$** | **Bias** | **Comments** |
|---|---|---|---|
| **y** | $(2 \times 10^{-28} + 9 \times 10^{-24} f^2)$ | $10^{-17}$ | See section 3.3. |
| **D$_\nu$** | $(4.5 \times 10^{-37} f^{-2} + 1 \times 10^{-28} + 9 \times 10^{-24} f^2)$ | $(10^{-17})$ | Bias determination limited by accelerometer and orbit modelling, $10^{-17}$ is clock limit |
| **a$_{NG}$** | $1.6 \times 10^{-18}$ (m/s$^2$)$^2$ | $5 \times 10^{-12}$ m/s$^2$ | |

**Tab. 2-1:** SAGAS uncertainties on science observables. Note that stated PSD are valid for integration down to the bias uncertainties. For longer integration, some further improvement on noise can be expected, but will be limited by the temporal variation of systematic effects, at presently unknown levels.

We will use the example mission profile elaborated in sect. 5. with a nominal mission lifetime of 15 years and the possibility of an extended mission to 20 years if instrument performance and operation allow this. In that time frame, the example trajectory allows the S/C to reach a heliocentric distance of 39 AU in nominal mission and 53 AU with extended duration. Most science objectives correspond to slowly varying effects over those timescales, so sampling rates of the measurements and corresponding required data transfer rates can be low. The only exceptions to this are the Post-Newtonian gravity test during occultation (sect. 2.3.) and the search for low frequency gravitational waves (sect. 2.8.). Even in those cases, 0.01 Hz sampling rates should be sufficient, which leads to tiny science data rates of three numbers / 100 s, amply accommodated by the optical link (kbps capacity, sect. 3.3.).

### 2.2. Test of the Gravitational Redshift and of Lorentz Invariance

The universal redshift of clocks when submitted to a gravitational potential is one of the key predictions of General Relativity (GR) and, more generally, of all metric theories of gravitation. It represents an aspect of the Einstein Equivalence Principle (EEP) often referred to as Local Position Invariance (LPI) [3]. In GR, the frequency difference of two ideal clocks is (to first order in the weak field approximation)

$$\frac{d\tau_S}{dt} - \frac{d\tau_G}{dt} \approx \frac{w_G - w_S}{c^2} + \frac{v_G^2 - v_S^2}{2c^2} + O(c^{-4}) \qquad (2\text{-}2)$$

with *w* the Newtonian gravitational potential at the location of the clocks and *v* their coordinate velocity. In theories different from GR the relation (2-2) is modified, leading to different time and space dependence of the frequency difference. This can be tested by comparing two clocks at distant locations (different values of *w* and *v*) via exchange of an electromagnetic signal. At present the most sensitive such experiment was carried out using two hydrogen maser clocks one on the ground and one on a parabolic orbit reaching up to 10000 km altitude [4]. It confirmed the GR prediction with a relative uncertainty of $7 \times 10^{-5}$. The SAGAS trajectory (large potential difference) and low uncertainty on the observable $y$ (directly the difference in 2-2) allows a relative uncertainty on the redshift determination given by the $10^{-17}$ bias on $y$ divided by the maximum value of $(w_G-w_S)/c^2$ reached. For nominal mission duration this corresponds to a test with a relative uncertainty of $1.0 \times 10^{-9}$ (with a very similar value for extended mission duration), which represents an improvement by a factor $7 \times 10^4$, almost 5 orders of magnitude. For a clear gravitational test it is necessary to measure the potential term in (2-2) independently of the velocity one ([4] in fact only measured the combination of both terms), *i.e.* one has to correct for the velocity term with the required $10^{-17}$ uncertainty. This implies that $\delta v_S/c$ needs to be determined to about $2 \times 10^{-13}$ ($v_S$ = 13 km/s at end of nominal mission), which should pose no difficulties given the noise level on the $D_\nu$ measurement (Tab. 2-1). Similarly, one has to correct for the potential from all Solar System objects (mainly the planets) with sufficient uncertainty. This should pose no particular difficulty either, provided the measurements are carried out when the S/C is sufficiently far from any massive object (see sect. 2.6. for details).

Tests of the universal gravitational redshift as described above can be viewed quite generally as comparing whether the gravitational potential governing the motion of freely falling test masses is the same as the one governing the evolution of two distant ideal clocks, *i.e.* whether a metric description of gravity is correct. A somewhat less general approach is to compare two co-located clocks of different type (so called null redshift tests) assuming that different types of atomic transitions are coupled differently to the ambient gravitational field (see also sect. 2.7. for a further interpretation in terms of variation of fundamental constants). The uncertainty of such null redshift tests is generally stated as the uncertainty with which the local frequency comparison limits a putative temporal variation $\delta y(t) = \delta w_{Sun}(t)/c^2$ varying at diurnal and annual frequency due to the rotation of the Earth and the eccentricity of the Earth's orbit. Presently, the best limit on null redshift tests is $1.4 \times 10^{-6}$, a factor 1400 less sensitive than the estimated SAGAS sensitivity. Even with further expected improvement of ground clocks SAGAS will conserve the advantage of the large variation of gravitational potential over the mission, about 30 times larger than attainable on the ground.

Additionally, SAGAS also provides the possibility of testing the velocity term in (2-2), which amounts to a test of Special Relativity (Ives Stilwell test), and thus of Lorentz invariance. Towards the end of the nominal mission, this term is about $4 \times 10^{-9}$ and can therefore be measured by SAGAS with $3 \times 10^{-9}$ relative uncertainty. The best present limit on this type of test is $2.2 \times 10^{-7}$ [6], so SAGAS will improve on present knowledge by a factor $\approx 70$. Considering a particular preferred frame, usually taken as the frame in which the 3K cosmic background radiation is isotropic, one can set an even more stringent limit. In that case a putative effect will be proportional to $(\mathbf{v}_S-\mathbf{v}_G) \cdot \mathbf{v}_{Sun}/c^2$ (cf. [6]), where $\mathbf{v}_{Sun}$ is the velocity of the Sun through the CMB frame ($\approx$ 350 km/s). Then SAGAS will allow a measurement with about $5 \times 10^{-11}$ relative uncertainty at best, which corresponds to more than 3 orders of improvement on the present limit. Such a scenario needs to be further studied to define the best suited S/C trajectory and estimate the resulting limit for that case.

Note that Ives-Stilwell experiments also provide the best present limit on a particularly elusive parameter ($\kappa_{tr}$) of the Lorentz violating Standard Model Extension (SME) photon sector [7], hence SAGAS also allows for the same factor 70 to $10^3$ improvement on that parameter.

## 2.3. Tests of Parameterised Post-Newtonian Gravity (PPN)

The PPN formalism describing a large class of metric theories of gravitation (including GR) in the weak field regime is well known (see *e.g.* [3]) and has been extensively tested in Solar System. The two most common parameters of the PPN framework are the Eddington parameters $\beta$ and $\gamma$, both equal to 1 in GR. With those two parameters the PPN metric to first PN order is

$$g_{00} = -1 + \frac{2w}{c^2} - \frac{2\beta w^2}{c^4}; \quad g_{0i} = -\frac{2(1+\gamma)}{c^3}w^i; \quad g_{ij} = \delta_{ij}\left(1 + \frac{2\gamma w}{c^2}\right) \tag{2-3}$$

where $w$ and $w^i$ are the scalar and vector potentials, to first approximation

$$w = \sum_A \frac{GM_A}{r_A}; \quad w^i = \sum_A \frac{GM_A v_A^i}{r_A} \tag{2-4}$$

with the sums carried out over all Solar System bodies.

The parameter $\beta$ appears only in $g_{00}$ and hence contributes to the equation of motion of test masses. For SAGAS the corresponding effect is a coordinate acceleration of $\approx 1.2 \times 10^{-10}$ m/s$^2$ at 1 AU and falling off quickly ($1/r^3$ dependence) as the satellite moves away from the Sun. Given the $5 \times 10^{-12}$ m/s$^2$ measurement uncertainty on $\mathbf{a}_{NG}$, which has to be corrected for in order to measure the gravitational motion, SAGAS is unlikely to set a limit on 1-$\beta$ better than about 4 %. This is not competitive with present limits ($\approx 10^{-3}$), so will not be further considered here.

The second parameter $\gamma$ has raised much theoretical attention. It characterises the amount of space-time curvature produced by unit rest-mass, and as such is affected by most types of modifications of GR. For example, it is modified at the low energy limit of string theory in certain cosmological models, which can lead to deviations from unity (GR value) as large as $10^{-7}$ to $10^{-5}$ [14], well within the range of SAGAS measurement (see below). Furthermore Scalar-Tensor models also lead to a deviation of $\gamma$ from unity [15] and may be related to dark-energy matter coupling [16], of interest in cosmology and fundamental physics.

Phenomenologically, the fact that $\gamma$ appears in the $g_{ij}$ and $g_{0i}$ terms of the metric leads to effects on light propagation and to gravito-magnetic effects. Present limits on $\gamma$ are obtained from measurements on light propagation (light deflection and Shapiro delay). The most stringent such limit was deduced from Doppler ranging to the Cassini mission during solar occultation (June 2002) yielding $\gamma = 1 + (2.1 \pm 2.3) \times 10^{-5}$ [17], in agreement with GR.

SAGAS will carry out measurements very similar to the Cassini one, during one or several solar conjunctions (depending on detailed trajectory) with improved sensitivity and at optical rather than radio frequencies, which significantly minimises effects from solar corona and the Earth's ionosphere. When the laser of the SAGAS link passes close to the Sun, the gravitational (Shapiro) delay leads to a modification of the Doppler observable $\delta D_\nu$ given by

$$\delta D_\nu(t) \approx \frac{d}{dt}\left[(1+\gamma)\frac{GM}{c^3}\ln\left(\frac{4 r_S r_G}{b^2}\right)\right] \approx -2(1+\gamma)\frac{GM}{c^3 b}\frac{db}{dt} \qquad (2\text{-}5)$$

where $b$ is the impact parameter (distance of closest approach) of the laser beam. At grazing incidence ($b \approx 7 \times 10^8$ m) and for a distant S/C (> 1 AU, then $db/dt \approx 30$ km/s; Earth orbital velocity) the maximum effect is about $8.5 \times 10^{-10}$.

When combining the on board and ground measurements such that the "up" and "down" signals coincide at the satellite (classical Doppler ranging type measurement) the noise from the on-board clock cancels to a large extent and one is left with noise from the accelerometer, the ground clock, and the atmosphere. We assume that for a twenty day measurement around occultation the accelerometer is operated in 1D along the direction of signal propagation (of interest here) leading to a factor of $\sqrt{3}$ improvement on its sensitivity given in sect. 3.1.2. and Tab. 2. We assume that ground optical clocks improve by about a factor 6 in stability with respect to best present performances to $\sigma_y(\tau) \approx 5 \times 10^{-16}/\sqrt{\tau}$ in terms of Allan variance (very likely by the time SAGAS is launched). Finally we assume atmospheric noise from turbulence and variations in temperature, pressure and humidity as described in sect. 3.3.4. Optimal filtering of the Shapiro delay signal in the total noise during a ten day measurement starting just after occultation leads to an uncertainty $\delta(\gamma) \leq 1.1 \times 10^{-7}$. Considering also the ten day measurement just before occultation allows another $\sqrt{2}$ gain, and $N$ occultations during the 20 year duration allow another gain of $\sqrt{N}$. However, observations are likely to be incomplete over the 20 days around occultation (data gaps due to loss of lock eg. from atmospheric fluctuations). Allowing for a factor $\approx 2$ loss (about a factor 4 loss in observation time) we thus conservatively estimate our overall uncertainty on the determination of $\gamma$ from a single occultation at $2 \times 10^{-7}$, with some potential for improvement with several occultations.

Furthermore, we have assumed that the non-gravitational accelerations of the S/C are simply measured and corrected without any modelling. However, it is likely that the non-gravitational S/C acceleration can be modelled over the short timescales involved during occultation to better than the accelerometer uncertainty, especially if occultation occurs when the S/C is far on its outward journey. In that case the measurement will ultimately be limited by the clock uncertainty rather than the accelerometer and measure most of the $8.5 \times 10^{-10}$ maximum amplitude of the effect, leading to an ultimate measurement uncertainty on $\gamma$ of $\leq 10^{-8}$.

The major systematic effect to be accounted for in Doppler measurements close to the Sun (apart from the non-gravitational acceleration discussed above) is the influence of solar corona, and stray light from the Sun "blinding" the telescopes. The latter is addressed in detail in Sect. 3.3. and found to be negligible even when pointing directly into the Sun. The dispersive nature of the solar corona leads to a time delay of electromagnetic signals proportional to $1/f^2$, the time variation of which shows up on the Doppler observable. In the Cassini experiment, this effect was measured and removed using Doppler observations at different frequencies (Ka and X band). The residual effect was estimated to be 4 orders of magnitude smaller than the measured effect, *i.e.* of order $10^{-14}$ [17]. For SAGAS at the optical frequency the full effect is expected to be about 8 orders of magnitude smaller ($1/f^2$ dependence) than for the Cassini Ka band. Additionally, it can be measured and removed using the combination of X-band and optical available on SAGAS, the large frequency difference allowing an even more precise measurement than with the Ka – X combination available on Cassini. So we are confident that solar corona effects will play no significant role and measurements down to solar grazing will be possible. In turn, it might be interesting to investigate the possibility of studying the solar corona using the precise measurements available during occultation, but for the time being this is not included as a scientific objective of SAGAS.

We note in passing that a similar test can be carried out during conjunction with Jupiter (trajectory allowing) with about 100 times less sensitivity but improved and different propagation systematics when grazing the planet. Combining the two would lead to significantly more confidence on the obtained bound or on the observation of a GR violation.

In summary, assuming only one occultation over the complete mission, we estimate that SAGAS will be able to measure the PPN parameter $\gamma$ with an uncertainty of about $2 \times 10^{-7}$, but more likely in the $10^{-8} - 10^{-9}$ region, when combining the accelerometer measurements with a model of non-gravitational accelerations over the short time of the occultation. These numbers represent an improvement by 2 to 4 orders of magnitude on best present results, and are well into the theoretically interesting $10^{-5} - 10^{-7}$ region and beyond [14], so should be able to significantly constrain present attempts at unification theories and cosmological models.

**2.4. Exploring Large Scale Gravity**
Experimental tests of gravity show a good agreement with General Relativity (GR) at scales ranging from the millimeter (laboratory experiments) to the size of planetary orbits. Meanwhile, most theoretical models aimed at inserting GR within the quantum framework predict observable modifications at smaller and/or larger scales.

Anomalies observed in the rotation curves of galaxies or in the relation between redshifts and luminosities of supernovae are ascribed to dark matter and dark energy components, the nature of which remains unknown. These dark components, which constitute 96% of the content of the Universe, have not been detected by non gravitational means to date. As the observed anomalies could also be consequences of modifications of GR at galactic or cosmological scales, it is extremely important to test the laws of gravity at the largest possible distances.

Such a test has been performed by Pioneer 10/11 probes during their extended missions. This largest scaled experimental test of gravity ever performed has failed to reproduce the expected variation of the gravity force with distance [18]. Precisely, the analysis of the radio-metric tracking data from the probes at distances between 20−70 astronomical units (AU) from the Sun has shown the presence of an anomalous, small, nearly constant Doppler shift drift, which can be interpreted as an unexpected acceleration of the order of $1 nm/s^2$, directed towards the Sun. The observation of this "Pioneer anomaly" has stimulated significant efforts to find explanations in terms of systematic effects on board the spacecraft or in its environment.

The inability to explain the anomalous behavior of the Pioneer spacecraft with conventional physics [18] has contributed to the growing discussion about its origin, a discussion which is still ongoing [41]. It has also motivated an interest in flying new probes to the distances where the anomaly was first discovered, that is, beyond the Saturn orbit, and studying gravity with modern techniques. A confirmation of the anomaly would constitute a breakthrough in fundamental physics, while a negative output would also be important, by considerably improving our knowledge of gravity laws at large distances.

Four letters of intent have been submitted to ESA after the Cosmic Vision call (DSGE, ODYSSEY, ZACUTO and SAGAS) with the aim of testing gravity laws in the outer Solar System. They were based on different measurement techniques and mission scenarios and also showed different levels for the mission objectives and technology readiness. The proponents have agreed to merge the four letters of intent to only two proposals, one L class (SAGAS) and one M class (ODYSSEY), combining and representing all four initial proposals. To that aim SAGAS has incorporated some of the ZACUTO science objectives, in particular the study of outer Solar System mass distributions (*e.g.* Kuiper belt) as detailed in sect. 2.5.

Over the past years, a large number of theoretical frameworks that allow for a scale (distance) dependent modification of GR have been suggested, *e.g.* generalized metric extensions of GR, Modified

Newtonian Dynamics, Tensor-Vector-Scalar-theory, Metric-Skew-Tensor Gravity, *f*(R) modified gravity theories, String theory and Cosmology motivated frameworks, Braneworld scenarios, and many others. It is far beyond the scope of the present proposal to even list all existing models, let alone describe and calculate the various observable effects one would expect for a mission like SAGAS. That type of activity has to be the subject of a dedicated theoretical study by a specific working group once the mission is at a more advanced stage and more details, in particular on mission profile and payload performance are known. However, it is very likely that for many of those theories SAGAS will contribute to closing an observational gap situated at distances covered by neither precision Earth based observation (*e.g.* LLR, *i.e.* Earth-moon distance) nor astronomical observations (> kpc).

We will restrict this section to a study of SAGAS in the context of large scale gravity, using the Pioneer anomaly (PA) as a quantitative example, and a few conventional and "new physics" hypotheses that may be used to explain it. In particular we will discuss some classes of conventional hypotheses considered in [18] and the two "extremes" of the generalized metric theory described in [19], a relatively large parameterized framework including some other models as special cases. The aim here is not so much to rigorously quantify the SAGAS measurements under the different hypotheses, but to show how the complementary SAGAS instruments allow the discrimination between the different hypotheses. Indeed, for tests of fundamental physics it is not only important to measure phenomena with highest accuracy, but also to address as many different aspects of a given theoretical approach as possible, thereby allowing a fine-tuning and cross-check between the different phenomenological consequences of a given theory.

We will consider the following classes of conventional hypothesis that could potentially cause observable effects similar to the PA, and discuss them in the context of SAGAS:

- **C1:** An insufficiently modeled non-gravitational acceleration
- **C2:** An additional Newtonian potential (*e.g.* Kuiper belt etc…)
- **C3:** An effect on the Doppler link that affects the Doppler observable $D_\nu$ (as on Pioneer) but not the frequency comparison observable *y* (*e.g.* a frequency shift in the signals on the trajectory or in the DSN antennae, that is identical for the up and down link, thus cancels in the difference used to measure the *y* observable).
- **C4:** An effect on the Doppler link that affects the radio signals of Pioneer but not the SAGAS optical link (*e.g.* unaccounted $1/f^2$ dispersion along the trajectory).

The general metric framework of [19] can be written in linearised form and to first approximation:

$$g_{00} = -1 + \frac{2w}{c^2} + 2\delta\Phi_N \ ; \quad g_{rr} = 1 + \frac{2w}{c^2} + 2\delta\Phi_N + 2\delta\Phi_P \quad (2\text{-}6)$$

where $\delta\Phi_N$ and $\delta\Phi_P$ are functions depending on *r* that need to be measured by experiment. In this framework the PA can be accommodated in two "extreme" cases corresponding to setting $\delta\Phi_P = 0$ or $\delta\Phi_N = 0$. Of course any intermediate combination of the two potentials can also be used, but here we will restrict our attention to the two extreme cases. Then the PA constraint implies

$$\delta\Phi_N = rl^{-1}; \quad \delta\Phi_P = 0 \qquad \text{"first sector"} \quad (2\text{-}7)$$

or

$$\delta\Phi_N = 0; \quad \delta\Phi_P = -\frac{c^2}{3GM}\frac{r^2}{l} \qquad \text{"second sector"} \quad (2\text{-}8)$$

where *l* is a characteristic length determined by the value of the observed PA to be $l \approx 10^{26}$ m. Note that the potentials need to take the forms described by (2-7) and (2-8) only at distances where the PA was observed (20 to 70 AU). This then leads to the two physics hypotheses that could cause potentially observable effects of the PA type on SAGAS:

- **P1:** The pure first sector of [19], equation (2-7)
- **P2:** The pure second sector of [19], equation (2-8)

To illustrate the versatility of SAGAS, we estimate the differences between the values of SAGAS observables under the different hypotheses on one hand, and when using the best fit orbital model from known physics, fitted to the Doppler observable (as done for Pioneer) on the other. The difference is estimated for a stretch of data covering one year when the S/C is at 30 AU with a velocity of ≈ 13.2 km/s. Tab. 2-2 shows the resulting differences.

| Hypothesis | $a_{NG}$ /m.s$^{-2}$ | $y$ | $D_\nu$ | Comments |
|---|---|---|---|---|
| C1 | 8.7x10$^{-10}$ | 4x10$^{-15}$ | 2x10$^{-10}$ | S/C closer and at lower $v$ than expected |
| C2 | - | 5x10$^{-14}$ | 2x10$^{-10}$ | Effect on $y$ mainly due to the additional grav. effect |
| C3 | - | - | 2x10$^{-10}$ | |
| C4 | - | - | - | No effect on SAGAS |
| P1 | - | 5x10$^{-14}$ | 2x10$^{-10}$ | Effect on $y$ mainly due to $\delta\Phi_N$ at 30 AU |
| P2 | - | - 9x10$^{-14}$ | 2x10$^{-10}$ | Effect on $y$ mainly due to $\delta\Phi_P$ $(v/c)^2$ at 30 AU |

**Tab. 2-2:** Anomalous effects on SAGAS observables under different hypothesis giving rise to PA type observations (see text for details).

The results shown in Tab. 2-2 are only rough estimates. A more detailed analysis, over the complete mission using simulated data needs to be carried out during the assessment phase to fully investigate the possibilities offered by SAGAS. Nonetheless, Tab. 2-2 allows two main conclusions:
1. With one year of integration, all SAGAS observables allow a measurement of any effect of the size of the PA with a relative uncertainty of better than 1% (taking into account the noise and bias uncertainties of Tab. 2-1).
2. The complementary observables available on SAGAS allow good discrimination between the different hypotheses, thereby not only measuring a putative effect, but also allowing a clean identification of its origin.

In summary, SAGAS offers the possibility to constrain a significant number of theoretical approaches to scale dependent modifications of GR. Given the complementary observables available on SAGAS the obtained measurements will provide a rich testing ground for such theories, with the potential for major discoveries that may well lead to a revolution of relativity and physics as a whole. In the light of present observational evidence at very large scales (galaxies, cosmology), and of interrogations as to the nature of dark matter and dark energy, experimental data at intermediate scales is much needed and SAGAS is well equipped to provide such information with uncertainties corresponding to the best of presently available technology.

### 2.5. Exploring Outer Solar System Masses

The exceptional sensitivity and versatility of SAGAS in the measurement of gravity can be used to study the sources of gravitational fields in the outer Solar System, and in particular the class of Trans Neptunian Objects (TNOs), of which those situated in the Kuiper belt have been the subject of intense interest and study over the last years [20]. Observation of Kuiper belt objects (KBOs) from the Earth is difficult due to their relatively small size and large distance, and estimates of their masses and distribution are accordingly inaccurate. Nonetheless, sufficient information is available to be able to confront it to models of formation of the Solar System, revealing some inconsistencies [20]. SAGAS will significantly contribute to the advancement of the knowledge on KBOs, thereby enlarging our knowledge on the Solar System and on the processes involved in its formation.

*The mass deficit problem in the Kuiper Belt :*
The Kuiper belt is the remnant of the circumsolar disk where the giant planets of the Solar System formed 4.6 billion years ago. Since 1992, more than 1000 Kuiper belt objects (KBOs) have been detected, mostly outside Neptune's orbit. The KBOs orbital elements revealed a complex structure. This structure is explained by perturbations by the planets, and in particular by resonances with Neptune. There is a large uncertainty in the disk mass due to conversion from absolute magnitude to sizes, assumptions about bulk density, ambiguities in the size distribution and strong limitation on the direct detection of small KBOs. Because of the steep size distribution, a large amount of mass can be in undetectable small KBOs. Estimates from the discovered objects range from 0.01 to 0.1 Earth masses, whereas in-situ formation of the observed KBOs would require 10 to 30 Earth masses of solid material in a dynamically cold disk. There are several hypotheses to explain this difference, a destruction of the distant planetesimal disk or a truncation of the original gas disk; these models would lead to low disk mass. Others models involve migration of the giant planets: the outwards migration of Neptune captures KBOs in migrating mean-motion resonances. This model would not perturb the outer disk, dynamically cold and undetectable by direct observations. In summary, large uncertainties on the total mass of the Kuiper belt, on its mass distribution and on masses of individual KBOs persist, and precise measurements of those quantities would significantly contribute to answering some of the questions related to these recently discovered Solar System objects and to the mechanisms of planet formation.

**2.5.1. Measuring the Kuiper Belt Mass Distribution**

The measurement sensitivity of a dedicated probe like SAGAS is of great interest for discriminating between different models for the spatial distribution of the Kuiper Belt. The most discussed models in literature are (see [21] for details and references):

- Two-ring models, which consists of two thin rings lying on the ecliptic with radius $R_1 = 39.4$ AU (resonance 3:2) and $R_2 = 47.8$ AU (resonance 2:1).
- The uniform disc model, which consists of a thin disc lying on the ecliptic, within distances $R_{Min} = 30$ AU and $R_{Max} = 55$ AU.
- Non-uniform disc, a thin disc with $R_{Min} = 30$ AU and $R_{Max} = 100$ AU and a mass function given by $f(r) = (r - R_{Min})^2/AU^2 \exp[-0.2(r - R_{Min})/AU]$.
- The toroidal mass distribution model, where mass is distributed in a toroid centered on the ecliptic with central radius $R_c = 42.5$ AU and thickness $R_t = 12.5$ AU.

One can plot the acceleration profiles for the different models [21] or, equivalently, the relative frequency shift due to the gravitational potential $\delta y = w_{KB}(r)/c^2$, as a function of heliocentric distance of the S/C, as shown in Fig. 2-1 [21].

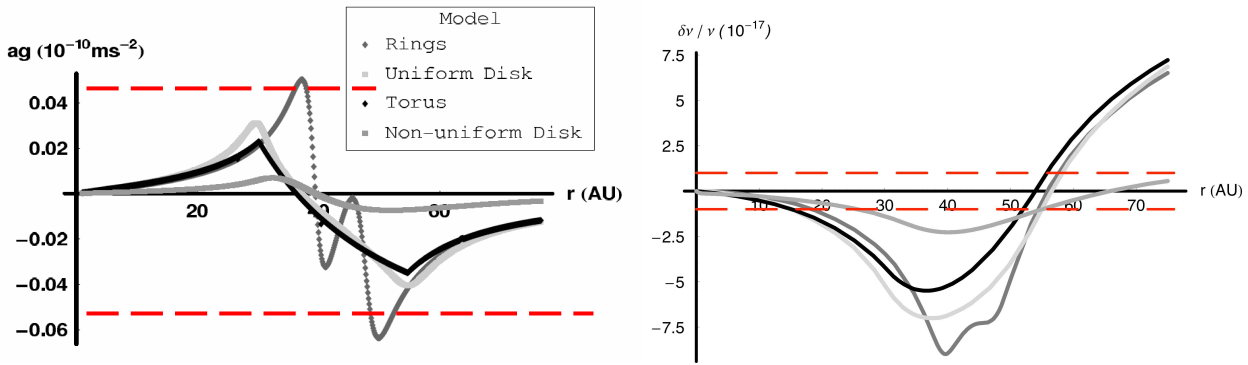

**Fig. 2-1:** Acceleration (left) and frequency shift (right) as a function of heliocentric distance from different Kuiper belt mass distributions (see text), with a total mass of $M_{KP} = 0.3\ M_E$ [21]. The red lines indicate the uncertainties of SAGAS.

Inspection shows that, apart from the weaker detectability of the non-uniform disk mass distribution, the remaining models can be essentially detected from distances beyond 15 AU, and well discriminated by the frequency difference observable at around 40 AU, ie towards the end of the nominal mission.

The $5 \times 10^{-12}$ m/s² uncertainty of the accelerometer on $\mathbf{a}_{NG}$ does not allow observation of the Kuiper belt gravity on the motion of the S/C. However, a complete analysis would involve modelling of the S/C trajectory and fitting the available data to that model. Calculating such models for different Kuiper belt mass distributions and fitting them to the total body of measurements ($\mathbf{a}_{NG}$, $y$, $D_v$) may well allow even better discrimination between the different distributions than suggested by Fig. 2-1 when using solely the frequency observable $y$.

A complete "scan" over all distances available during the mission not only allows the determination of the shape of the curves shown on Fig. 2-1 and hence the mass distribution, but also the amplitude *i.e.* the total mass $M_{KP}$. The accuracy of that will obviously depend on the distribution. For example, in the "two rings" distribution SAGAS will determine $M_{KP}$ with an uncertainty of at least 10% *i.e.* about 0.03 Earth masses. In a more detailed analysis one would fit all measurements during the mission to the candidate curve thereby likely decreasing the overall uncertainty by at most (depending on correlations of individual measurements) $\sqrt{N}$ where N is the number of measurements. Given that the clock noise integrates to $10^{-17}$ in about 10 days and for the ≈ 14 years travel from 15 AU to the end of the extended mission (53 AU), this is another potential factor 22 improvement.

We note that the gravity of the planets yields potentially much larger signals than the Kuiper belt gravity, however these depend strongly on distance and their effect can be taken into account when designing the trajectory (see sect. 2.6.).

**2.5.2. Study and Discovery of Individual Kuiper Belt Objects**
All known KBOs have been discovered using Earth based observations and/or the Hubble space telescope [22] and, as mentioned above, the weak signals, conversion from absolute magnitude to sizes, assumptions about bulk density etc… lead to large uncertainties on their masses and mass distributions. As shown, the SAGAS frequency observable $y$ is well suited to study the large, diffuse, statistical mass distribution of KBOs essentially due to its sensitivity directly to the gravitational potential ($1/r$ dependence), rather than the acceleration ($1/r^2$ dependence). That large diffuse signal masks any signal from individual KBOs rendering measurements of the individual properties difficult. The situation is modified when closely approaching one of the objects. Indeed, the crossover between the acceleration sensitivity (given by the $5 \times 10^{-12}$ m/s$^2$ $a_{NG}$ uncertainty) and the frequency sensitivity ($10^{-17}$ uncertainty on $GM/(rc^2)$) for an individual object is situated at about 1.2 AU. Below that distance, the acceleration measurement is more sensitive than the frequency one. This suggests a procedure to study individual objects using the SAGAS observables: use the S/C trajectory (corrected for $a_{NG}$) to study the gravity from a close object and subtract the diffuse background from all other KBOs using the frequency measurement. Tab. 2-3 below lists some of the known KBOs that are within the reach of SAGAS, and the uncertainty with which their mass can be determined using the SAGAS observables when approaching to 0.5 AU or to 0.2 AU.

| Object | Semi major axis / AU | Estimated Mass/$10^{21}$ kg | $\delta M/M$ @ 0.5 AU | $\delta M/M$ @ 0.2 AU |
|---|---|---|---|---|
| Pluto | 39.5 | 13.05 | 0.03 | 0.005 |
| (136108) 2003 EL$_{61}$ | 43.3 | 4 | 0.1 | 0.02 |
| (136472) 2005 FY$_9$ | 45.8 | 4 | 0.1 | 0.02 |
| Quaoar | 43.4 | 2 | 0.2 | 0.03 |
| Ixion | 39.7 | 0.6 | 0.7 | 0.1 |

**Tab. 2-3:** Some KBOs and the uncertainty with which SAGAS will be able to measure their mass.

Given that there are no particular constraints on the SAGAS trajectory for the other science objectives (apart from the occultation required for the PPN test), it should be possible to choose a trajectory that leads SAGAS close to a KBO to study (to be investigated in a more detailed mission scenario). This is even more likely given the rate of discovery of KBOs over the last years, so it is plausible that many more options for KBO flybys will exist by the time SAGAS is launched. Finally, we mention the possibility of discovery of one or several KBOs by SAGAS itself as it flies by them, again a more detailed study is required to determine the optimal trajectory in order to maximise the probability of discovery.

In summary, because of its complementary observables SAGAS offers the unique possibility of exploring the statistical distribution, the total mass as well as individual objects of the Kuiper belt and thereby significantly enhance our knowledge of this largely unknown part of our Solar System.

**2.6. Knowledge of Planetary Gravity**
The gravitation from planets and other large bodies of the Solar System is likely to cause systematic shifts in the SAGAS measurements when the S/C is close to them. These shifts can be corrected for using present knowledge of planetary gravity. Tab. 2-4 shows the critical distance $r_C$ from each planet below which the corrections for the planetary effect can no longer be calculated to the required accuracy for SAGAS. In other words SAGAS can no longer achieve its science objectives when at a distance less than $r_C$ from the planet. In turn, the SAGAS observables then provide information on planetary gravity. For example the Jupiter fly-by, (closest approach ≈ 600000 km) will allow a measurement of $GM_{Jup.}$ potentially at $1.4 \times 10^{-11}$ in relative value when limited by the $5 \times 10^{-12}$ m/s$^2$ uncertainty of the accelerometer. However, a more detailed study is needed to determine how well the instruments and the Doppler ranging will operate in close vicinity to Jupiter, so for the time being we adopt a more conservative estimate of $\leq 10^{-10}$ for that measurement. Even at that accuracy (about 100 fold improvement on present knowledge) ample information will be collected concerning not only $GM$ but also higher order terms of the potential and Jupiter's moons, providing detailed information on the planet and its moons. In case of a 2$^{nd}$ fly-by to another planet, similar measurements could be performed for that planet as well. Note that such measurements remain complementary to the other science objectives, as the time spent below $r_C$ is generally short compared to the total mission duration (< 50 days for Jupiter).

| Planet  | $\delta GM/GM$ | $r_C$ /AU | Reference for $\delta GM/GM$ |
|---------|----------------|-----------|------------------------------|
| Jupiter | 2x10$^{-8}$    | 0.15      | R.A. Jacobson, JUP230 orbit solution, (2003) |
| Saturn  | 3x10$^{-8}$    | 0.1       | R.A. Jacobson, AJ **132**, 2520, (2006) |
| Uranus  | 2x10$^{-6}$    | 0.3       | Yoder, C. F., in ed. T. J. Ahrens, *Global Earth Physics: A Handbook of Physical Constants*, American Geophysical Union, Washington DC, (1995) |
| Neptune | 2x10$^{-6}$    | 0.4       | |

**Tab. 2-4:** Present relative uncertainties on planetary gravitational constants, and critical distance for SAGAS (see text)

## 2.7. Variation of Fundamental Constants

Spatial and/or temporal variations of fundamental constants constitute another violation of LPI and thus of GR. Over the past few years, there has been great interest in that possibility (see e.g. [8] for a review), spurred on the one hand by models for unification theories of the fundamental interactions where such variations appear quite naturally, and on the other hand by recent observational claims of a variation of different constants over cosmological timescales [9, 10]. Such variations can be searched for with atomic clocks, as the involved transition frequencies depend on combinations of fundamental constants and in particular, for the optical transition of the SAGAS clock, on the fine structure constant $\alpha$.

More generally, such tests take two forms: searches for a drift in time of fundamental constants, or for a variation of fundamental constants with ambient gravitational field. The latter tests for a non-universal coupling between ambient gravity and non-gravitational interactions (clearly excluded by the EEP) and is well measured by SAGAS, because of the large change in gravitational potential during the mission.

For example, changing parameters of the standard model are usually associated with the effect of massless (strictly speaking, very light) scalar fields. One candidate, much discussed in the literature, is the dilaton, which appears in string models. Other scalars naturally appear in string-theory inspired cosmological models, in which our Universe is a "brane" floating in a space of larger dimensions. Such scalar fields would couple to ordinary matter and thus their non-zero value would introduce a variation of fundamental constants, in particular $\alpha$ of interest here. The non-zero value of such scalar fields could be of cosmological origin [12,13], leading to a constant drift in time of fundamental constants, and/or of local origin, *i.e.* taking ordinary matter as its source [11]. In the latter case one would observe a variation of fundamental constants with the change in local gravitational potential, which can be parameterized in the simple form [11]

$$\frac{\delta \alpha}{\alpha} = k_\alpha \, \delta\!\left(\frac{GM}{rc^2}\right). \tag{2-2}$$

The best present limit on $k_\alpha$ is obtained in [11] from a comparison between a Hg$^+$ optical and Cs microwave clock, which have a sensitivity to the variation of $\alpha$ of -3.2 and +2.8 respectively. Monitoring their relative frequency as a function of the changing solar potential on the Earth's surface (varying by $\delta(GM/(rc^2)) \approx 3.3 \, 10^{-10}$ due to the Earth's eccentricity), a limit of $k_\alpha < 6\times 10^{-7}$ was obtained.

The difference in gravitational potential between the Earth and the SAGAS satellite at the end of nominal mission is about $\delta(GM/(rc^2)) \approx 9.7\times 10^{-9}$, which is 30 times more than the variation attainable on Earth. The Sr$^+$ optical transition used in the SAGAS clock has a sensitivity to the variation of $\alpha$ of $\approx 0.43$. When compared to a ground clock with 10$^{-17}$ uncertainty, this yields a limit of $k_\alpha < 2.4\times 10^{-9}$, a factor 250 improvement over the best present limit.

Note that, even with expected improvement of ground clocks, SAGAS will always keep the advantage of the large variation of gravitational field which is not attainable on the ground. On the other hand, the relatively low sensitivity of the Sr$^+$ transition to a variation of $\alpha$ (0.43 as compared to the 3.2 for Hg$^+$ for example) is a disadvantage in this respect. Using one of the other candidate ion species for SAGAS (see sect. 3.2.) would reduce that disadvantage (*e.g.* 0.88 sensitivity for Yb$^+$).

## 2.8. Upper Limits on Low Frequency Gravitational Waves

Doppler ranging to deep space missions provides the best upper limits available at present on gravitational waves (GW) with frequencies of order $c/L$ where $L$ is the S/C to ground distance *i.e.* in the 10$^{-3}$ to 10$^{-5}$ Hz range [23, 25], and even down to < 10$^{-6}$ Hz, albeit with lower sensitivity [24, 25]. The corresponding limits on GW are determined by the noise PSD of the Doppler ranging to the spacecraft for stochastic GW backgrounds [24, 25], filtered by the bandwidth of the observations when looking for GW with known signatures *e.g.* sinusoidal GW from binaries [23, 25]. In the former case best limits [25] are about $10^{-13}/\sqrt{Hz}$ in GW strain sensitivity around 0.3 mHz and, in the latter case, about $h \leq 2\times10^{-15}$ for the maximum amplitude of sinusoidal GW, again at 0.3 mHz. These limits increase rapidly at lower frequency [25].

In the case of SAGAS, using optimal data combination allows a strain sensitivity of $\sim 10^{-14}/\sqrt{Hz}$ for stochastic sources in the range of $10^{-5}$ to $10^{-3}$ Hz [42], limited at low frequency by the accelerometer noise that could possibly be improved by modelling of non-gravitational accelerations. When searching for GW with particular signatures in the $10^{-5}$ to $10^{-3}$ Hz frequency region, optimal filtering using a corresponding GW template will allow reaching strain sensitivities as low as $h \approx 10^{-18}$ with one year of data, or even lower if more data is available. This corresponds to three orders of magnitude improvement on best present limits.

In spite of these very low limits, it is at present considered unlikely that known sources could generate GW of that amplitude in the corresponding frequency region. The stochastic cosmological GW background (with PSD falling off as $f^{-1.5}$ in most cosmological models) is already constrained to levels below the sensitivity of SAGAS by Pulsar observations, albeit in the nHz frequency region. To obtain useful information on that it would be necessary to extend the frequency range to lower frequencies, down to $10^{-6}$ or $10^{-7}$ Hz, which requires good modelling of low frequency non-gravitational motion of the S/C. Potentially more interesting could be non-stochastic sources like, for example, inspiraling and merging Black Hole Binaries (BHB). However, even those are expected to provide signals with amplitudes around $10^{-19}$ or less, so about a factor 10 outside the visibility of SAGAS.

In summary, SAGAS will improve on best present upper limits on GW in the $10^{-5}$ to $10^{-3}$ Hz frequency range by about four orders of magnitude. Although it is not expected at present that GW with sufficiently large amplitudes can be found in that region, the obtained results might still be useful as upper bounds for astrophysical models of known GW sources, whilst leaving open the door for potential surprises.

**2.9. Technology Developement**
SAGAS technology choices are based on cold atom and laser technology, both particularly adapted for tracking, timing and communication over large distances and measurement of DC (or very slowly varying) effects because of the absolute reference provided by the atoms. The mission takes advantage from important heritage on cold atom technology used in ACES/PHARAO and laser link technology designed for LISA. It will provide an excellent opportunity to develop those technologies for general use in interplanetary missions, including development of the ground segment (DSN telescopes and optical clocks) that will allow such technologies to be used in many other mission configurations for precise timing, navigation and broadband data transfer throughout the Solar System.

Of particular interest in this respect is the synergy between the different payload elements. The atoms in the accelerometer provide the absolute frequency reference for the quartz USO necessary for operation of the optical clock. The optical clock, in turn, provides the ultra-narrow and accurate laser (locked to the atoms) which makes operation of the link in minimal configuration possible. The complete payload thus provides an optimal ensemble for the implementation of such technologies in an operational mode, ready to be re-used in future deep-space, planetary, or terrestrial missions.

# 3. PAYLOAD

The SAGAS payload is composed of two instruments (accelerometer and optical clock) and the optical link which is used as a scientific instrument (frequency comparison and Doppler measurements), and also for data transfer. The particular feature of the SAGAS payload is the synergy between the different payload parts which significantly simplifies the overall design and reduces cost and technology development: The accelerometer also provides the absolute calibration with respect to the Cs atoms of the quartz USO, which in turn is used to generate all on-board RF signals used in the optical clock, the link and the time-tagging of observations. The optical clock and optical link share the same frequency, thereby avoiding use of a femtosecond frequency comb and allowing common use of some of the laser sources, with the clock providing a highly narrow and stable laser for the link. In short, the combination of on board instruments is not only optimal and complementary for the science objectives, but also close to ideal for the simplicity and coherence of the technology.

## 3.1. Cold Atom Accelerometer

### 3.1.1. Introduction

To reach the scientific objectives of SAGAS, accurate measurements of accelerations along three orthogonal axes are required. The design payload and its characteristics arise from the development of the fields of cold atom physics and atom interferometry [40]. Key technologies are identical to those already developed within the ACES project for the PHARAO payload. Compared to PHARAO, the SAGAS accelerometer shares identical key technologies and similar payload architecture and subsystems. Concerning issues related to atom interferometry, it also benefits from the different studies carried out within the HYPER project. Ground developments of cold atom interferometers have already shown performances comparable to state of the art optical interferometers [26]. Measurement of the gravity acceleration is limited on Earth to a few parts in $10^9$ by environment effects: tides, atmospheric pressure and underground water fluctuations..., which vanish in a space environment. The intrinsic accuracy of cold atom interferometers makes them attractive for studies in fundamental physics [27], *e.g.* for the determination of the gravitational constant *G* [28] or for the measurement of the Planck constant $\hbar$ [29] (either by recoil measurement, or via the watt balance experiment) and for applications in the fields of geophysics and geodesy.

Atom interferometers can perform measurements with very high sensitivities, taking advantage of the absence of gravity and of a very low vibration environment, which allows to significantly increase the interrogation time. A high level of accuracy can be reached, using cold atoms, thanks to an excellent control of the atomic trajectories and the atom-laser interaction, as is the case in atomic clocks. In addition, this payload can operate as a micro-wave clock to control the drift of the ultra-stable quartz oscillator at a $10^{-12}$ level independently from the optical clock, and thereby provide a stable and accurate on-board reference time scale.

### 3.1.2. Principle of operation and baseline choice

The accelerometer is based on the use of cold atoms and Raman transitions for the manipulation of the atomic wave-packets. The atoms are alkaline atoms, which can be easily cooled using all solid-state semi-conductor diode lasers. The Raman transitions couple the two ground states of the alkaline atoms (noted $|g\rangle$ and $|e\rangle$) and can be realized by the same lasers as the cooling. The two Raman lasers are propagating in opposite directions and transfer a momentum $\hbar k$ to diffracted atoms (corresponding to a velocity of the order of 1cm.s$^{-1}$). The phase shift due to acceleration is given by: $\Delta\phi = -\vec{a}.\vec{k}.T^2$. The sensitivity depends only on the wave vector *k* and the square of the time between pulses *T*. The three axes of acceleration are successively measured within the same vacuum tube using three orthogonal pairs of Raman lasers.

For each measurement, the sequence (total duration $T_c \approx 3$ s) of preparation of the atomic sample, interferometer and detection is similar and follows this order:
- Cooling of the atomic sample: (duration *790 ms*)
    - Loading of the atoms in the molasses (high power and low detuning of cooling lasers)
    - Cooling of the atoms to 1 µK temperature in state $|e\rangle$ (reduction of power and increase of the detuning of the cooling lasers)
- Preparation of the input state: (duration *10* ms)
    - Increase of the bias magnetic field to degenerate the magnetic sub levels
    - Micro-wave pulse to transfer the atoms in the $|e, M_f=0\rangle$ state to the $|g, M_f=0\rangle$ state ($M_f=0$ states are less sensitive to magnetic field)
- Discarding the remaining atoms in states $|e, M_f \neq 0\rangle$ using a pusher laser beam
- Velocity selection Raman pulse: transfer of the atoms to $|e, M_f=0, P = -\hbar k/2\rangle$: pulse of about 50 µs
- Discarding of the remaining atoms in states $|g, M_f=0\rangle$

Interferometer based on three Raman pulses (called π/2, π, π/2): (duration 2s)
- First pulse to split the initial atomic wave-packet in a coherent superposition of two partial wave-packets with different momenta: $|e, M_f=0, P = -\hbar k/2\rangle$ and $|g, M_f=0, P = +\hbar k/2\rangle$
- Second pulse (after a time *T* = 1s) to exchange the two states and redirect the two partial wave-packets
- Third pulse (after another interval *T*) to recombine them when they overlap.

Detection: determination of the transition probability (duration *200 ms*)
- Measurement of the number of atoms in state $|e\rangle$
- Descarding of the atoms in state $|e\rangle$

- Re-pumping of the atoms in $|g\rangle$ to $|e\rangle$ and measurement of the number initially in state $|g\rangle$

### 3.1.3. Choice of Raman transitions for the interferometer
The beam splitters of the interferometer can be based on Raman transition or on Bragg transitions. In this second case, the atoms are diffracted with two photon transitions but without change of internal state. The advantages of this method are linked to the fact that atomic wave packets stay in the same internal state, which strongly reduces a number of perturbing effects (laser phase noise, light shift, magnetic field fluctuations, collisional shift,…). The problem is that it needs Raman transitions in any case for the velocity selection and for the detection as it uses the difference of internal states (a Raman transition is then used to change the internal state of one of the two output ports). As most interferometers studied so far use Raman transitions, we keep this as the baseline of the payload, but further studies will be carried out to validate this strategy. In any case, the two options can be interchanged without any physical change in the payload, but by simply changing the pulse sequence, which can be done by software only.

### 3.1.3. Choice of the atoms
In principle any alkaline atom can be chosen for the atomic source. The intrinsic sensitivity of the interferometer is very similar for all alkaline atoms as it depends only on the wave vector $k$ and the interaction time $T$ but not on the mass of the atom. In practice, technical differences have to be taken into account to optimize the sensitivity or accuracy:
- Only Cs, Rb, and K can easily be cooled using laser diodes
- TRL is better for Cs thanks to developments for the space project PHARAO/ACES
- The total volume of the interferometer decreases as $m^3$, where $m$ is the mass of the atom considered, for a given interrogation time as: (i) the temperature of the atomic sample decreases with the mass, thereby reducing the velocity dispersion, and (ii) the maximal splitting between the atomic wave-packets decreases with the mass, thereby reducing the physical length of the atom interferometer. The reduction of the size limits the requirement on the gradient of gravity induced by the satellite
- Collisions between cold atoms give rise to differential shifts between the two partial wave-packets leading to a bias on the acceleration signal

The Caesium atom is the best choice for all points except for the collisional shift, for which $^{87}$Rb has much lower shift (two orders of magnitude). This last point will be addressed in the next paragraph.

### 3.1.4. Choice of molasses as atomic source
Different possible sources of cold atoms can be used for the atom interferometer: molasses, magneto-optical trap (MOT), ultra-cold sources from evaporative cooling in a magnetic or an optical trap (in degenerate state or not). The choice of an optical molasses has been driven by the simplicity, lower mass and power consumption but will give reduced performances.

Compared to ultra-cold sources, the residual temperature (1µK for Cs) gives an extension of the atomic cloud of about 1.5 cm/s (FWHM), which limits the total interrogation time to a few seconds and so the sensitivity per unit of time. But, as explained below, this sensitivity is still good enough to achieve the performance required for SAGAS. Compared to a MOT, the molasses avoids the need of magnetic coils for the trapping, which dissipates ≈ 10 W. The main disadvantage is the reduction of the number of captured atoms. Nevertheless, the number of useful atoms at the end of the sequence will be of the order of $3\times10^5$ and is not a limiting point of the experiment. Another advantage is the drastic reduction of the collisional shift due to cold atom interactions during the interferometer measurement, as the initial size is much bigger (reduction by two orders of magnitude), allowing the use of Cs atoms. The calculation of this shift, for $10^6$ atoms at 1µK with a typical initial size of 15 mm (FWHM) and a total interrogation time $2T=2$s, gives a shift of $3\times10^{-13}$ m.s$^{-2}$, negligible compared to the accuracy needed for SAGAS.

The use of the 3D MOT or the use of a 2D MOT, to load the molasses, can be considered as an option to increase the number of cold atoms (more than one order of magnitude). In the two cases 10 W are needed for optimum loading. But a compromise can be found concerning the gradient of magnetic field: for example with a twice smaller current (dissipation of the order of 2 W) the captured atoms are still almost 10 times higher than in molasses.

### 3.1.5. Acceleration sensitivity
The sensitivity to acceleration per cycle is given by $1/(kT^2.\text{SNR})$, where SNR is the signal to noise ratio of the measurement. This sensitivity improves with the averaging time $t$ as $1/\sqrt{t}$. Different sources of noise limit the SNR: atom shot noise, phase noise between Raman lasers, acceleration noises.

***Shot noise:*** A fundamental limit is given by the atom shot noise due to the finite number of atom (N) leading to a quantum projection noise during the detection process. The maximum SNR is then proportional to $\sqrt{N}$, which corresponds to 500 in the present case.

***Raman frequency reference noise:*** A technical limit is coming from the residual phase noise between Raman lasers. Ground studies have shown that noise contribution from phase lock loops between lasers can be sufficiently reduced to remain well below the phase noise due to the imperfections of the reference frequency at 9.2 GHz. By using the spectral noise density of the PHARAO flight model frequency reference, we obtain a SNR=200 per shot (3s total) leading to a sensitivity limit of $3 \times 10^{-10}$ m.s$^{-2}$ per shot. As the sensitivity improves with the square root of the measurement time, the required sensitivity ($5 \times 10^{-12}$ m.s$^{-2}$) is obtained after only 3 h per axis, which gives 9 h for the three axes. One should mention than specific optimization of the frequency reference for atom interferometers should further reduce this contribution. This noise source is presently the limit to the sensitivity of the accelerometer.

***Microvibration from fly wheels:*** Limitations may also come from microvibrations of the S/C caused by unbalance in the fly wheels used for pointing stability. In the case of vibration at frequencies $f \gg 1/T$, the sensitivity is decreasing as $1/(2\pi f)^2$, giving typically five orders of magnitude reduction for frequencies around 100 Hz. Using vibration amplitudes from Bepi-Colombo mission studies (1 N at a frequency of $f = 66$ Hz) with a S/C mass of 1000 kg, the modulation of the acceleration signal seen at low frequencies by the aliasing effect is $2.3 \times 10^{-8}$ m.s$^{-2}$ in the worst case ($f = (2n+1)/(2T)$). As this modulation is periodic and the key feature is the averaging of the acceleration during long term, this term becomes negligible compared to phase noise limitations for averaging times longer than 3 hours. It can be further reduced by choosing the cycle time $T_C$ such that the noise frequency is not a harmonic of the cycling frequency $1/T_C$. Moreover, as the sensitivity to high frequency acceleration noise presents some zero at harmonics of $1/T \approx 1$s (best case), a fine-tuning of interrogation time to the wheel frequency allows a very good cancellation of these residual terms making them negligible.

### 3.1.6. Acceleration accuracy

To achieve the requirement for SAGAS, the accelerometer has to be accurate at a level of $5 \times 10^{-12}$ m.s$^{-2}$. Compared to ground experiment, the absence of gravity, and so of average velocity of the atomic cloud allows reduction of most of the systematics. Moreover, some of the error sources decrease with increasing interaction time $T$, as they depend on the interaction with the Raman laser. These considerations make possible the extrapolation from ground performances to needs for SAGAS. Main sources of systematic error are:

***Magnetic field:*** The difference of magnetic shifts of the two ground states of the Cs atoms gives in general a difference of phase shift at the output of the interferometer. In fact, the sensitivity to a bias magnetic field is zero in this kind of symmetric interferometer, as the two atomic wave packets spend the same time in the two internal states. Moreover, as the average velocity of the atomic wave packet is zero, it is also not sensitive to gradients of magnetic field, which is not the case in ground experiments.

***One photon light shift from Raman lasers:*** To cancel the differential light shift between the two states induced by the Raman lasers, the ratio of power between Raman lasers has to be adjusted (typical ratio is 1.8). If the ratios are not perfect but equal for the first and the last pulse, this effect is cancelled at first order. Only residual effects due to the expansion of the atomic cloud during the sequence, and to the gradient of intensity on the Raman lasers lead to a phase shift. For an average intensity change of 10% between the first and the last pulse, the ration between Raman powers has to be kept at 0.35% to achieve the required accuracy. Moreover, as it has been shown on ground, this shift can be cancelled to $10^{-3}$ by alternative measurements with opposite directions of diffraction ($\pm \hbar k$) making this effect negligible.

***Two photon light shift from Raman lasers:*** To limit effects from wave front distortion (see paragraph below), we will use co-propagating Raman lasers retro-reflected by a mirror. The presence of four Raman beams in the interferometer zone allows for two diffraction processes in opposite directions, degenerated by the Doppler effect due to the recoil velocity. The forbidden transition is not very far off resonance and induces a differential shift between the two partial wavepackets, which does not cancel by the beam reversal technique. Again, the atomic phase shift is only a residual effect due to the expansion of the atomic cloud between the first and last pulse. If the average intensity seen by the atom differs by 10% between the first and last pulse, the intensities of the Raman lasers has to be controlled to 0.7% to achieve the required accuracy. The measurement of the bias due to this effect can be done during the initial mission phase (when no science measurements are required) by extrapolation to zero power by alternating measurements with different Raman powers, and correcting for the effect during the rest of the mission, with additional periodic calibrations if necessary.

*Coriolis effect:* This point is critical for ground experiments due to the rotation of the Earth, but is negligible in SAGAS.

*Laser wave front distortions:* The acceleration is measured with respect to the wave fronts of the lasers, which have to be well controlled. As the atoms interact with the same part of the lasers for the three pulses, this effect is cancelled at first order. The residual effect is due to the expansion of the atomic cloud between the first and the last pulse. The use of state of the art reflection mirrors ($\lambda/1000$ over 3 cm diameter) and a calibration procedure (extrapolation to zero with a factor 50), by changing the interaction time and/or the temperature of the atomic sample, allows keeping this source of bias under expected accuracy. An in depth calibration campaign could be carried out during early mission phase with regular later checks as required. This effect is the most critical in terms of accuracy and should be assessed in detailed ground studies at an early stage.

To achieve the required performances the accelerometer needs a well control environment:

*Requirements on vibration noise:* Microvibrations from the SC have to be kept small to not degrade the performances of the accelerometer, especially at frequencies that are harmonics of the cycling frequency $1/T_C$. Indeed, as the measurement has dead time, the aliasing effect of frequencies close to the harmonics of $1/T_C$ appears as limits to the sensitivity. In case of white acceleration noise, the level has to be lower than $7 \times 10^{-10}$ m.s$^{-2}$.Hz$^{-1/2}$. We expect the main contribution to come from vibrations due to the fly-wheels of the AOCS, which was found to be negligible (see sect. 3.1.5. above).

*Requirement on self gravity:* Self gravity from the SC may bias the accelerometer, as for any accelerometer. This acceleration has to be known to better than the expected accuracy: $5 \times 10^{-12}$ m.s$^{-2}$, which requires placing the accelerometer at the S/C centre of mass and a uniform distribution of S/C mass. These points need to be addressed in the detailed S/C design (see also sect. 4.2.)

### 3.1.7. Use as micro-wave clock

As all the elements needed to realize a micro-wave atomic clock at moderate accuracy ($10^{-12}$ in relative frequency) are present in the payload, one can measure periodically the drift of the reference quartz oscillator, which is used to generate all radio or micro-wave frequencies for the accelerometer, the optical clock, the optical link and the on-board time scale. The atomic source and detection system are the same as for the accelerometer and the interrogation may be done by the micro-wave antenna used for the state preparation. Sensitivity of $10^{-12}$ per shot and the required $10^{-12}$ accuracy can easily be achieved.

### 3.1.8. Description of the payload

Most of the subsystems and components are similar or can be derived from the PHARAO project [30]. Estimations of mass and power budget take advantage of these similarities. The atomic accelerometer is composed of four subsystems (see fig. 3.1.-1): the vacuum tube where the atoms are cooled and interrogated (essentially the capture-zone sub-system of PHARAO), the optical bench for cooling and Raman transitions, the frequency reference for state preparation and Raman reference and the on board management unit to control the complete sequence and the acquisition.

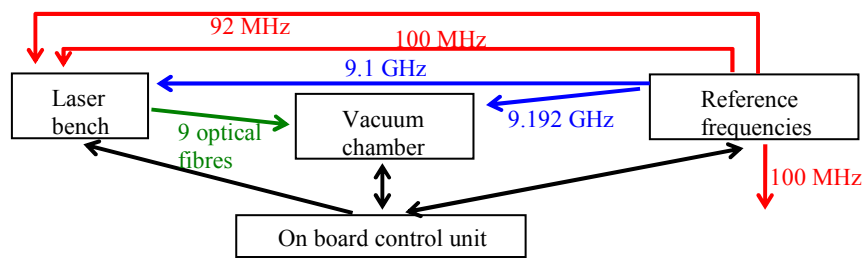

**Fig. 3.1.-1:** Atomic accelerometer subsystems

The **vacuum tube** is the central part of the payload where the atomic sample is prepared and the interferometer takes place. The vacuum has to be kept at $10^{-8}$ Pa level. It has to be non magnetic and protected again outside stray magnetic fields using a magnetic shield. It is mostly realized in titanium to fulfil the requirements on the magnetic field, support mechanical constraints during the launch, and be as light as possible. It is composed of:
- a Caesium oven to control the Cs pressure thanks to a valve aperture and a temperature control
- vacuum pumps: getter pump and ion pump (outside the magnetic shield)

- a central chamber with 6 cooling beams, 3 Raman laser pairs for the 3 acceleration measurements, 2 light detection collection systems
- a micro-wave antenna for the state selection and for clock measurements
- a two layer magnetic shield to control stray magnetic fields
- three pairs of coils for bias magnetic fields to raise the degeneracy of magnetic sub levels

The mass budget is 21 kg (16 kg for the magnetic shield) and a power budget of 5 W.

The **optical bench** provides lasers beams for the cooling, the interferometer and the detection, with well controlled frequencies and powers (see fig. 3.1.-2). As cooling beams and Raman beams are used at different time and have very close frequencies, the same lasers generate them [31]. Again key technologies benefit from development done for the PHARAO project. They are based on extended cavity lasers at 852 nm using interference filters for the selection of the wavelength and piezo-electric actuators for the fine tuning. Three extended cavity lasers are needed at the same time:

- the first (ECL1), frequency locks on a Cs cell as reference laser
- the second (ECL2) frequency locks by comparison with ECL1 and serves as repumper laser during the cooling and the detection and reference Raman laser during the interferometer phase
- the third (ECL3) frequency locks by comparison with ECL2 for cooling and detection and phase locks by comparison with ECL2 during the Raman phase.

The frequencies of these latter two lasers are controlled by mixing the optical beat signal (in the range of 8.5 to 9.2 GHz) with the microwave reference at 9.1 GHz, via frequency to voltage converter (FVC) or phase lock loop [31]. The two lasers (ECL2 and ECL3) are amplified thanks to a tapered amplifier. Again, using a common amplifier for the two lasers reduces complexity. Four acousto-optic modulators (AOM) are used to control the powers and to switch from cooling beams to Raman beams: one for the cooling and one for each Raman direction. The use of AOM avoids the need of mechanical actuators, which may age faster. The power consumption is quite small, as the three AOM for Raman pulses are switched off almost all the time and the driven frequency is the same. Six mechanical shutters are needed to avoid any stray light from cooling beams during Raman phase and vice-versa. Nine polarization-maintaining fibres are used to send the six cooling beams and the three Raman beams to the vacuum chamber. The same beams (with different frequencies) are used for the Raman transitions and the detection. Taking into account the duration of the mission, triple redundancy is planned for each laser. Switching from one laser to a spare one is done thanks to a rotating half wave plate and a polarizing cube. The mass budget is 20 kg and the power budget 25 W (including electronics).

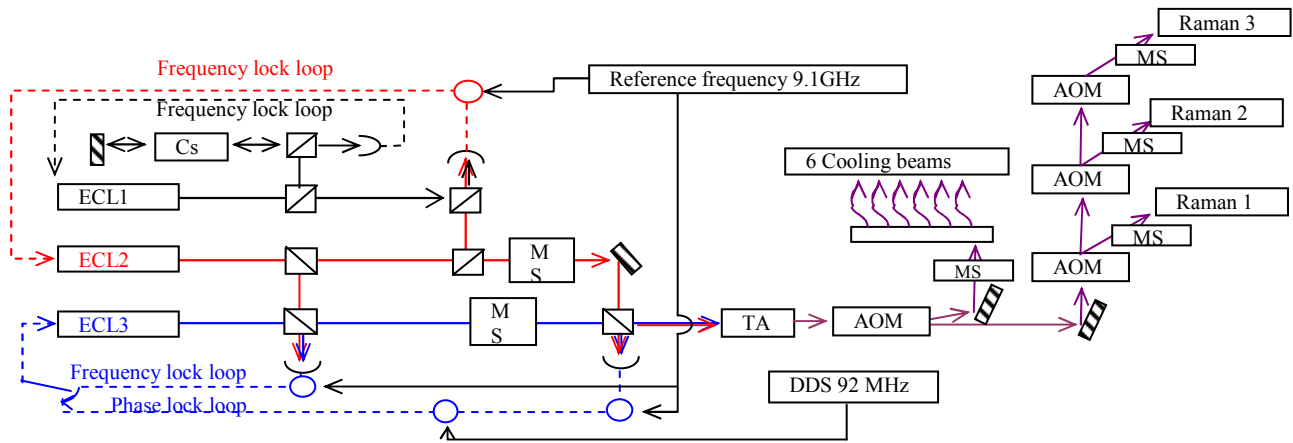

**Fig. 3.1.-2:** Optical bench subsystem (ECL=Extended Cavity Laser, Cs=Cs vapour cell, MS=Mechanical Shutter, TA=Tapered Amplifier, AOM=Acousto Optic Modulator, DDS=Direct Digital Synthesiser).

The **frequency reference** provides frequencies for the Raman transition, the state selection (and clock function) and AOM drivers (100 MHz) and frequency lock loops of lasers. It also provides references in radio frequency range for other payloads: ion clock and optical link. The design is identical to the one of PHARAO, giving the same performances in terms of frequency stability (below $10^{-13}$ from 1 to 10s) and phase noise [30]. It is based on an ultra stable oscillator at 5 MHz and a 100 MHz intermediate quartz oscillator (phase locked to the 5 MHz) multiplied to 9.1 GHz. The 9.192 GHz signal for state selection (and or clock mode) is realized thanks to a final mixing with a radio-frequency signal at 92 MHz provided by a direct digital synthesis (DDS). The 9.1 GHz signal is also used as reference for the frequency lock loops of ECL2 and 3 and for the phase lock loop of ECL3. In this last case, the radio frequency signal derived from the comparison of the optical beating and the 9.1 GHz reference is then used for phase lock loop of ECL3 by comparison with a 92 MHz signal from the DDS mentioned previously. For the frequency reference, the mass budget is 7 kg and the power budget is 12 W.

The **On board Control Unit** allows realization of the measurement sequence, control of the frequency and power of the lasers, data acquisition and analysis. One can use PHARAO heritage as the requirements are similar. The mass budget is 6 kg and the power budget 26 W.

| **Subsytem** | **Mass** / kg | **Power** / W<br>DC-DC converter included | **Dimensions** / mm |
|---|---|---|---|
| Vacuum tube | 21 | 5 | 400x400x500 |
| Optical bench + electronics | 20 | 25 | 500x300x200 |
| Reference frequencies | 7 | 12 | 300x300x100 |
| On board unit control | 6 | 26 | 250x250x120 |
| Total | 54 | 68 | 125 litres |

**Table 3.1.-1:** Cold atom accelerometer mass, power, volume budgets

### 3.1.9. Current heritage and Technology Readiness Level (TRL)
Demonstrations of DC accelerometer on the vertical axis (gravimeter at $\leq 10^{-8}$ m.s$^{-2}$), limited by the presence of gravity and vibrations, have been done on ground, thus validating the method. The technology used in the payload is the same as PHARAO in the ACES project: same lasers, similar vacuum chamber, same frequency references. The engineering model of PHARAO passed all tests. We thus estimate the technology readiness at level 6.

## 3.2. Optical Trapped Ion clock

### 3.2.1. Clock specifications
The clock component of the overall mission scenario is based on the use of an on-board optical frequency standard with frequency stability $\leq 10^{-17}$ for 10 day integration times (ie $\sigma_y(\tau) \leq 1\times10^{-14} \tau^{-1/2}$), where clock frequency data is downlinked to Earth by means of a high power link laser. This required performance is one order of magnitude better than the PHARAO clock and requires the development of a new generation of space clocks. Instead of the microwave frequencies used in PHARAO, the SAGAS clock uses optical frequencies. Ground based optical clocks are evolving at a significantly fast rate, with one or two single ion clocks already demonstrating instabilities ~ $4\times10^{-17}$ at $10^4$ s and $2\times10^{-17}$ accuracy in realising the unperturbed ion frequency [32]. There is good reason to expect similar performance for a range of other optical clock systems, which is extendable to below the target specification at longer times. The range of options available to satisfy these stability / reproducibility requirements is briefly outlined below.

### 3.2.2. Choices for optical clocks
Currently, there are two distinct high accuracy optical atomic clock architectures, one based on a single cold trapped ion approach, and one on cold neutral atoms held within an optical lattice. The single ion architecture presently represents a higher technology readiness level than is the case with neutral atom lattice clocks. In addition neutral atom clocks exhibit larger mass, volume and power consumption when compared to the ion clock arrangement. As a result, only the single ion clock is considered for this SAGAS proposal.

Within this ion clock arena, there are a number of possible ion species and isotopes which have been developed with state-of-the art characteristics. These are itemised in Table 3.2.-1 below. On examination, it can be seen that quantum-limited theoretical stabilities for all species surpasses the stability requirement of $10^{-17}$ @ 10 days. The limited amount of experimental stability data to long averaging times in excess of $10^3$ s existing (eg for $^{199}$Hg$^+$, $^{27}$Al$^+$ and the $^{171}$Yb$^+$ quadrupole clock transition at 435 nm), exhibits instability performance a factor 3 to 5 above this limit, but still within the $10^{-17}$ @ 10 days specification.

| Ion | Clock Transition | λ nm | Life-time s | Natural linewidth | Theor. stability @ 100 s | Recorded stability @ 100 s | Theor. stability @ $10^6$ s (~ 10 days) |
|---|---|---|---|---|---|---|---|
| $^{199}$Hg$^+$ | $^2S_{1/2} - {}^2D_{5/2}$ | 282 | 0.09 | 1.3 Hz | $1.3 \times 10^{-16}$ | $3 \times 10^{-16}$ | $1.3 \times 10^{-18}$ |
| $^{171}$Yb$^+$ | $^2S_{1/2} - {}^2D_{3/2}$ | 435 | 0.05 | 3.1 Hz | $2.7 \times 10^{-16}$ | $9 \times 10^{-16}$ | $2.7 \times 10^{-18}$ |
| $^{88}$Sr$^+$ | $^2S_{1/2} - {}^2D_{5/2}$ | 674 | 0.4 | 0.4 Hz | $1.5 \times 10^{-16}$ | $1.3 \times 10^{-15}$ | $1.5 \times 10^{-18}$ |
| $^{40}$Ca$^+$ | $^2S_{1/2} - {}^2D_{5/2}$ | 729 | 1 | 0.15 Hz | $1.0 \times 10^{-16}$ |  | $1.0 \times 10^{-18}$ |
| $^{115}$In$^+$ | $^1S_0 - {}^3P_0$ | 236 | 0.2 | 0.8 Hz | $0.7 \times 10^{-16}$ | $5 \times 10^{-15}$ | $0.7 \times 10^{-18}$ |
| $^{171}$Yb$^+$ | $^2S_{1/2} - {}^2F_{7/2}$ | 467 | $2 \times 10^6$ | ~ nHz | $2.8 \times 10^{-17}$ |  | $0.3 \times 10^{-18}$ |
| $^{27}$Al$^+$ | $^1S_0 - {}^3P_0$ | 267 | 21 | 8 mHz | $1.6 \times 10^{-17}$ | $3 \times 10^{-16}$ | $0.2 \times 10^{-18}$ |

**Table 3.2.-1:** *Theoretical stability assumes interrogation times equal to lifetime, except for $^{171}$Yb$^+$ octupole 467 nm transition and the $^{27}$Al$^+$ 267 nm transition, where 5 s interrogation time is assumed in both cases.*

In respect of these quantum limited stabilities, the absence of the quadrupole shift, the small black body shift and the small magnetic field sensitivity, the *J*=0-0 clock transitions could be considered better choices for a high-specification clock system with excellent stability and low systematics. However, for continued long-term performance within the space mission environment, the complexity of the particular trap technology and the difficulty to generate deep UV wavelengths for cooling or clock transitions are also strong determinants. These UV wavelengths are normally generated from fundamental frequencies in the mid IR, which then require to be doubled and doubled again, often with the doubling crystals housed within power enhancement cavities in order to generate the requisite level of power. Additionally, the need for UV optics and UV fibre-based delivery systems are also likely to increase complexity, with such fibre and coatings suffering enhanced radiation damage within the space environment.

These, and other considerations, preclude the selection of the $^{199}$Hg$^+$ and $^{27}$Al$^+$ clocks. Particularly, the $^{199}$Hg$^+$ clock requires UV clock and cooling transitions, as well as being cryogenically cooled. The $^{27}$Al$^+$ clock has a UV clock transition, and requires the use of a quantum logic algorithm to read out the clock transition. Additionally, the $^{115}$In$^+$ clock requires both UV cooling and clock transitions, with quadrupling of IR high power diode and solid state lasers respectively. The $^{171}$Yb$^+$ $^2S_{1/2} - {}^2F_{7/2}$ 467 nm clock transition is very weak, and requires tight tolerances and special alignment arrangements for spatial overlap with the ion. As a result, these systems are not considered feasible for a long term space mission.

### 3.2.3. Detailed comparison of viable ion clock choices for SAGAS

With the rejection of these complex ion clock species, three possibilities remain, with optical wavelengths for the clock transitions, and optical or near UV for cooling. Table 3.2.-2 compares the magnitudes of the most significant physics parameters that will contribute to potential frequency shift and uncertainty for these choices, and Table 3.2.-3 shows a comparison of laser technology issues.

| Ion | Clock Transition | λ nm | Electric quadrupole moment ($ea_0^2$) of the excited state | Blackbody Stark Shift ($10^{-16}$) at room temperature |
|---|---|---|---|---|
| $^{171}$Yb$^+$ | $^2S_{1/2} - {}^2D_{3/2}$ | 435 | 2.08(11) | 5.8 |
| $^{88}$Sr$^+$ | $^2S_{1/2} - {}^2D_{5/2}$ | 674 | 2.6(3) | 6.7 |
| $^{40}$Ca$^+$ | $^2S_{1/2} - {}^2D_{5/2}$ | 729 | 1.83 | 9.7 |

**Tab. 3.2.-2**: *Comparison of significant physics parameters contributing to systematic frequency shifts for the 3 ion clock options*

Examination of Table 3.2.-2 shows there is little difference in magnitude for the electric quadrupole moment of the upper state (which gives rise to the quadrupole shift) or the blackbody shift for the 3 options. As a result, the maturity of laser technology needed to generate the various cooling, auxiliary and clock wavelengths is compared to determine the feasibility of the three choices.

| Ion | Clock λ nm | Clock laser technology | Cooling λ nm | Cooling laser technology | Technical complexity |
|---|---|---|---|---|---|
| $^{171}$Yb$^+$ | 436 | SHG of 872 nm ECDL diode laser | 369 | SHG of 738 nm ECDL diode laser | Doubling stages for both clock & cooler |
| $^{88}$Sr$^+$ | 674 | ECDL diode laser | 422 | SHG of 844 nm ECDL diode laser | Doubling stage for cooler |
| $^{40}$Ca$^+$ | 729 | ECDL diode laser | 397 | SHG of 794 nm ECDL diode or UV ECDL diode | Doubling stage or UV diode for cooler, difficult clock wavelength |

**Tab. 3.2.-3:** *Options for SAGAS ion clock considered feasible; SHG second harmonic generation, ECDL extended cavity diode laser*

The technology issues for each option are examined in turn. The $^{171}$Yb$^+$ clock requires doubling stages for both clock and cooler. The fundamental wavelengths are not served well with available tapered amplifiers at this time (though this could change), with the cooling fundamental wavelength at 738 nm being a particularly difficult region from this point of view. Additionally, there is the requirement for the clock laser wavelength (or harmonic) to provide the link laser wavelength with ~ 1 W of power. Again, no significant power at the clock wavelength is currently available.

The $^{88}$Sr$^+$ clock requires only one doubling stage from 844 nm to 422 nm for laser cooling. Sufficiently powerful tapered amplifiers exist with a few hundred mW output such that relatively simple single pass doubling in periodically poled KTP can provide enough 422 nm power to provide all three cooling/compensation beams. Narrow-linewidth probe lasers at the 674 nm clock transition are readily available with extended cavity diode lasers. Further, high power 674 nm tapered amplifier lasers are now available at the 500 mW level, providing good opportunity for its use also as the link laser wavelength.

Extended cavity diode lasers for probing the $^{40}$Ca$^+$ ion clock transition at 729 nm are available (eg Toptica Photonics AG), but it is unclear at this time the extent of the linewidth reduction available from these commercial devices, or the ease with which the bare diodes can be obtained. On the other hand, high power tapered amplifiers exist at the clock wavelength, offering opportunity for the link laser. Cooling of the ion at 397 nm can be achieved by doubling 794 nm, and like $^{88}$Sr$^+$, tapered amplifiers exist so that single pass doubling should be possible.

In summary, it can be seen that possibilities exists for all these three options. However, taking together the currently available power levels for both clock and cooler, the need for only one single pass doubling stage, clock laser linewidth and the clock accuracy already achieved, it is considered that the $^{88}$Sr$^+$ 674 nm clock represents the most feasible ion clock option at this time. It is acknowledged that the technology underpinning all three options has the capability to evolve in the near to medium term, and this should be a core component of technology refinement activity during early stages of the L class mission preparations.

### 3.2.5. $^{88}Sr^+$ ion clock system

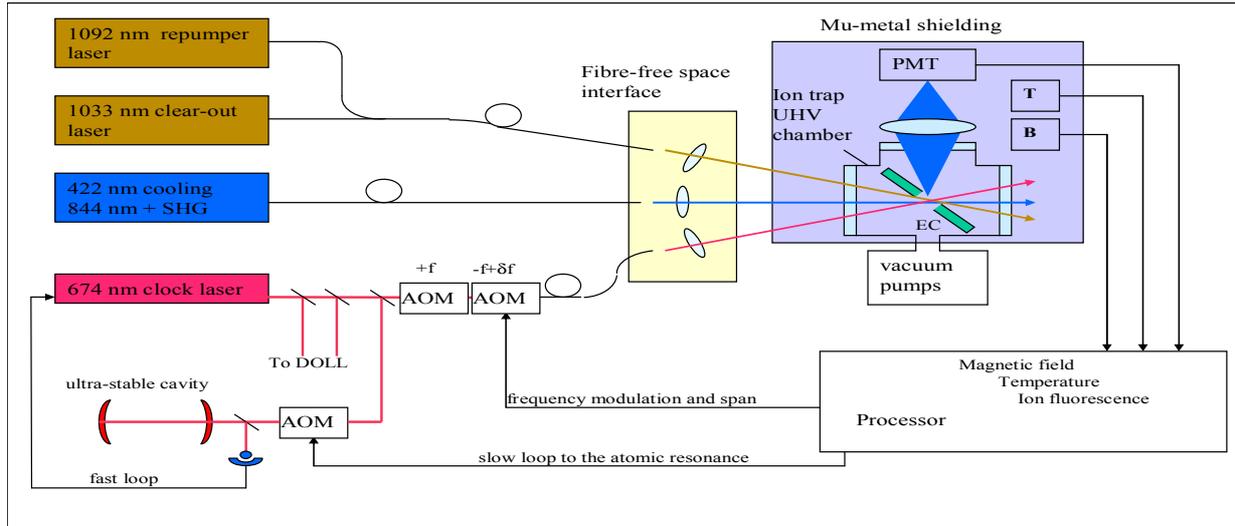

**Fig 3.3.-1:** *$^{88}Sr^+$ ion clock system architecture: AOM, acousto-optic frequency shifter; PMT, photo-multiplier; SHG, second harmonic generation; EC, end-cap trap; T, temperature sensor; B, magnetic field sensor. Redundant laser units not shown, nor control lines for active unit monitoring and redundant unit activation after major fault diagnosis, nor standard monitoring and control lines.*

The ion clock system design is outlined in figure 3.3.-1. It comprises a number of sub-components, including:

- the primary physics package, an RF end-cap trap for ionising and confining a single $^{88}Sr^+$ ion within an ultra-high vacuum chamber pumped by a small ion pump and non-evaporable getter pump. External magnetic field coils in 3 orthogonal axes allow the nulling of external fields, and setting of a fixed field. These coils are surrounded by mu-metal shielding to minimise external field changes. While three dimensional single ion traps with electrode separations in the range 0.5 mm - 1.5 mm are currently used for frequency measurements and proposed here, we will also investigate the possibility to benefit from novel segmented ion trap technologies with integrated optics.
- a laser platform to provide Doppler cooling of the ion to ~ 1 mK on the strongly-allowed $^{88}Sr^+$ ion dipole transition at 422 nm. This is achieved by frequency doubling of ~ 200 mW commercial extended cavity diode laser at 844 nm, by means of second harmonic generation (SHG) in periodically poled KTP.
- auxilliary lasers for (1) repumping the ion from the $^2D_{3/2}$ metastable level during the cooling sequence, and (2) for fast clear-out of the clock transition metastable $^2D_{5/2}$ upper level, once the clock transition has been driven. These are temperature- and current-stabilised DFB lasers.
- the 674 nm clock laser probing the $^2S_{1/2}$ - $^2D_{5/2}$ quadrupole clock transition. This is an extended cavity diode laser frequency FM-stabilised to a very high finesse ultra-low-expansion (ULE) cavity mounted on a temperature-stabilised and evacuated platform.
- a high NA lens imaging and photomultiplier detection system to record the statistics of 422 nm fluorescence quantum jumps as a function of 674 nm clock laser frequency, providing the cold ion linewidth reference which steers the ULE cavity-stabilised clock laser light.
- a fibre system to deliver the cooling, auxiliary and clock light from source to trap, making use of achromatic doublets where necessary at the fibre-free space interface for launching into the trap
- a monitoring and control processor driving the clock sequence (cooling-probing-detection). The processor also monitors frequency and amplitude data necessary to determine normal laser and ion operational conditions and initiate resetting and recovery algorithms where necessary, and laser unit failure.
- a redundancy level of 2 or 3 units for both cooling, clock and high power link laser, plus a redundancy level of 3 units for the repumper and clear-out DFB lasers. All redundancy units for each wavelength to be fibre multiplexed as standard, allowing redundant unit activation on determination of prior unit failure mode.

### 3.2.6. $^{88}Sr^+$ ion clock performance and critical issues
The progress in the accuracy of single ion optical clocks has been rapid and it is foreseeable that within the next few years the level of 1 part in $10^{17}$ will be demonstrated in several ground based systems, including $Sr^+$. For some time, the shift due to the interaction of electric field gradients with the D-levels used in the quadrupole reference transitions in Hg$^+$, Yb$^+$ and Sr$^+$ was considered as the most critical systematic. Now, it was established that without active compensation, this shift is generally smaller than 10$^{-15}$ and that compensation schemes can suppress it by more than two orders of magnitude. Since these schemes rely on an averaging over several Zeeman components, the temporal stability of the magnetic field (static value: about 1 µT) becomes a critical issue. Mu-metal shielding of the trap vacuum system will be employed in order to limit drifts of the magnitude of the magnetic field to a value below ten nT per hour.

The second critical issue is the interaction of the ion with blackbody radiation emitted from the trap. The sensitivity cofficient is d(ln $f$) / d$T$= 3x10$^{-25}$ $T^3$ / K$^4$ so that for operation at 300 K, d$T$ should be less than 2 K. If temperature changes are slow, no active stabilization will be necessary. Characterisation of the thermal environment and measurement of the temperature with thermistors will be sufficient. Another critical point related to temperature issues is the laser pointing alignment which should be kept within typically ten micrometers.

Reloading of the trap will be necessary over a mission period of 15 years. A miniature dispenser, containing a few mg of Sr and a hot wire as an electron source are needed. Automatic reloading of the trap will induce an interruption of clock operation for a few minutes. Charging problems of the trap and vacuum system will be avoided by illumination with ultraviolet LEDs. The discharge procedure will also be used periodically to eliminate charges from high energy radiation

The $^{88}$Sr$^+$ ion clock systematic frequency dependencies are shown in table 3.2.-4, in the form of the expected uncertainty budget. Overall uncertainty of ~10$^{-17}$ is achievable within the spacecraft environment, provided external magnetic field and temperature variations are sufficiently low or adequately controlled.

| Influence | Coefficient/ condition | Bias | Uncertainty | Comment |
|---|---|---|---|---|
| **Magnetic field** <br> • Linear Zeeman shift <br> • 2$^{nd}$ Order Zeeman shift | Applied field ~ 1 µT <br> ±5.6 Hz /nT (Δm=0) <br> 5 µHz/µT$^2$ (Δm=0) | 0* <br> 6 µHz | ~ 10$^{-17}$ <br> <<10$^{-17}$ | Zeeman pair averaging + mu-metal shielding |
| **Electric field** <br> • Quadrupole shift <br> • Low freq AC Stark <br> • Clock laser AC Stark | 3 Zeeman pair average <br> 3D micromotion nulled <br> 0.5 mHz / Wm$^{-2}$ | 0** <br> 0 <br> 150 µHz | ≤ 10$^{-17}$ <br> < 10$^{-17}$ <br> < 10$^{-18}$ | See ** <br> <br> 30 nW in 300 µm |
| **Temperature** (± 1 K) <br> • Blackbody shift | 4 mHz/K (room temp) | 300 mHz | ~ 2x10$^{-17}$ | Assumes ± 1 K, BB coefficient uncertainty large but fixed |
| **2$^{nd}$ Order Doppler shift** <br> • Resid. thermal motion <br> • Residual micromotion | T ~ 1 mK <br> 3D micromotion nulled | 0 <br> 0 | ~ 10$^{-18}$ <br> ~ 10$^{-18}$ | |
| * Continuous Zeeman component pair averaging on 20 s cycle time removes 1$^{st}$ order Zeeman effect;     Residual due to B field drift rate at trap with 1-layer mu-metal shield < 0.8 nT / min sufficient for averaging down, but 2-layers (~ 1 nT / hour) better for contingency, dependent on external field variation encountered. <br> ** Magnetic field stability / directionality of ≤ 10 nT /hour by 1- or 2-layer mu-metal shielding, dependent on external field variation encountered. ||||| 

**Tab. 3.2.-4:** *$^{88}$Sr$^+$ ion clock systematic frequency shift dependencies*

### 3.2.7. Clock payload requirements
Payload requirements are developed by extrapolation from existing ground-based clock arrangements, together with attention to existing laser and opto-electronic hardware and reference to the existing ACES/PHARAO microwave clock payload. This includes a redundancy level of three lasers for both the clock laser ECDL and the high power 844 nm cooling laser and single pass doubler. The use of available DFB lasers for the 1092 nm repumper laser and 1033 nm clear-out laser will also allow a redundancy level of at least three per wavelength.

| Type of optical clock | power (W) | Physics volume | Electronics volume | Physics mass | Electronics mass | Total mass |
|---|---|---|---|---|---|---|
| Ion clock | 80 | 150 litres | 30 litres | 50 kg | 30 kg | 80 kg |

**Tab.3.2.-5:** *Projected space ion clock power, volume and mass*

The physics package volume of 150 litres is considered an upper limit. It will include all laser systems, opto-electronic beam conditioning and fibre launching and delivery to the trap package, beam manipulation onto the ion, photomultiplier detection of the ion fluorescence, trap vacuum chamber and 2 ls$^{-1}$ ion pump plus non-evaporable getter pump, 3-axis magnetic field coils for field nulling and quantisation axis definition, plus mu-metal shielding. The clock laser system will include one high finesse supercavity maintained within a small vacuum housing and pumped by a mini ion pump. The cooling/auxiliary lasers will require lower finesse smaller reference cavities for controlled wavelength tuning and stabilisation.

Not considered in the above are the general requirements for environmental temperature control and radiation shielding. For the former, encapsulation of the science payload within a temperature-controlled spacecraft interior should help to maintain free space optical alignments. For the latter, a certain level of lead shielding may be necessary to take account of the varying radiation environments encountered.

### 3.3. Deep space Optical Laser Link (DOLL)

We propose an original optical link concept for SAGAS that takes full advantage of the particular technology available on-board and the synergy between payload components (narrow, stable and accurate laser from the clock, accurate microwave from the accelerometer) whilst being specially tailored to achieve the required science objectives. Concerning the technology, particular emphasis was paid to making maximum use of existing developments (ACES/PHARAO diode laser technology, LISA telescope technology, SLR/LLR ground stations,). DOLL features in particular:
- continuous wave laser operation in both directions (two-way system).
- heterodyne detection schemes on-board and on ground.
- high data transfer rate with simultaneous science measurements.
- asynchronous operation allowing optimal combination of on-board and ground measurements.
- large stray light rejection from heterodyne detection and due to the possibility of large controlled frequency offset between the up and down link using the accurate on-board microwave.

More generally, we believe that DOLL will not only find its use in SAGAS, but more generally contribute significantly toward the development of optical, high accuracy interplanetary navigation and broadband communication.

#### 3.3.1. Principle of operation and estimated performance
DOLL is based on continuous two-way laser signals exchanged between the ground station and the S/C, with independent heterodyne detection of the incoming signal at either end (no transponder scheme). The fundamental measurement is the frequency difference between the local oscillator and the incoming signal. This measurement is particularly adapted to SAGAS because of the availability of the very narrow clock laser ($\approx$ 10 Hz linewidth) on board and on the ground, stabilized to the atomic transitions with a stability of $\sigma_y(\tau) = 1 \times 10^{-14} \ \tau^{-1/2}$. The on-board and ground data are combined in post treatment and analysed in order to extract the science observables: Doppler ($\rightarrow$ velocity difference), clock frequency difference, ranging.

This allows asynchronous operation, *i.e.* combining the measurements taken at different times in order to independently optimise each observable by maximum rejection of error sources for each observable (see [42] for details). For example, the clock frequency difference observable is obtained by differencing the S/C and ground measurements. This rejects all frequency shifting effects that are common to the up and down link (Doppler, atmosphere, etc…) up to path asymmetries between the up and down link which need to be corrected for. In general it is possible to choose the measurements that are combined in such a way as to optimise that cancellation. For example, differencing the S/C measurement taken at $t_0$ and the ground measurement taken at $t_0$-$D/c$ (with $D$ the S/C to ground distance) *i.e.* combining the signal that arrived at the Earth at $t_0$ with the one that left the Earth at the same instant, rejects most of the atmosphere effects. Similarly, the Doppler observable is obtained by adding the up and down link, so summing, the S/C measurement taken at $t_0$ and the ground measurement taken at $t_0$+$D/c$, *i.e.* combining the signal that arrived at the S/C at $t_0$ with the one that left the S/C at the same instant rejects the S/C clock instability. The level to which such cancellation is effective, is determined by how accurately one can time S/C and ground measurements, estimated for SAGAS to be about 10 ns (see section 3.3.5.).

To avoid complexity onboard the spacecraft, the signals emitted on the ground will be offset in frequency to largely compensate for the Doppler frequency shift (up to 70 GHz), so the frequency received at the S/C is close to the nominal clock frequency, allowing direct heterodyne beat with the local clock laser.

Additionally, the up and down signals will be linearly polarized with orthogonal polarizations (for stray light rejection), which implies variable polarization direction for the ground station telescope.

Table 3-1 below summarizes the performance of DOLL for the main observables, when using optimal rejection of noise sources in asynchronous operation (see also section 3.3.4).

| Observable | Performance | | Remarks |
|---|---|---|---|
| | Noise | Bias | |
| Doppler ($D_\nu$) | $S(f) = (1 \times 10^{-28} + 4.3 \times 10^{-23} f^2)$ Hz$^{-1}$ | < $10^{-17}$ | Clock and Troposphere limited |
| Frequency diff. ($y$) | $S(f) < 10^{-28}$ Hz$^{-1}$ | << $10^{-17}$ | Well below clock performance |
| Ranging ($r$) | 4 km$^2$ Hz$^{-1}$ | < 3 m | FSK or PSK modulation at 1 kHz |
| Data transfer | ≈ 3000 bps | | Satellite at 30 AU |

**Tab. 3-1:** Summary of DOLL performance. Ranging and data transfer values are @ 30 AU.

Table 3-1 does not take into account accelerometer noise. If one wants to extract the purely gravitational trajectory of the S/C from the Doppler data the noise on the non-gravitational acceleration measured by the accelerometer needs to be added (cf. Tab. 2-1). Similarly, the frequency difference needs to be corrected for Doppler and relativistic effects again adding accelerometer noise or troposphere noise depending on how the up-link and down-link measurements are combined (cf. Tab. 2-1).

Note that for orbit determination the Doppler and ranging observables are, in principle, redundant. For SAGAS the precise orbit is obtained from the Doppler, whilst the ranging (orders of magnitude less precise) serves only to improve initial conditions for the integration.

### 3.3.2. Space segment

The main subsystems of the DOLL space segment are the telescope and the optical bench providing the laser source.

The present baseline is to use a telescope design similar to that of the LISA mission, adapted for the SAGAS wavelength (674 nm) and including a high definition CCD camera (similar to COROT or LISA). The LISA telescope consists of a Cassegrain telescope characterized by an aperture diameter of 400 mm, a system focal length of 4800 mm, and a comparatively large magnification of 80. Main reflector M1 and subreflector M2 are separated by 450 mm with the help of a CFRP tube spacer. The telescope ocular includes a pupil of 5 mm diameter in which a "Point-Ahead Angle Mechanism" can be placed for the correction of the out-of-plane point-ahead angle. In the present SAGAS baseline design, both the transmitted and the received signals are processed with the same telescope. A two telescope system (one for emission, one for reception) may be envisaged as an alternative if stray light from the emitted signal turns out to be a limiting problem for the detection of the received signal, but present estimations indicate that this will not be necessary (see section 3.3.4.2.).

The optical bench houses a high power laser (LH) for the link, a low power laser (LL) for the clock, and an ultra stable cavity for short term laser stability (≈ 10 Hz linewidth), the long term stability being achieved by locking to the atoms in the optical clock (see fig. 3-1). At present semiconductor laser systems (ECDL) with tapered amplifiers that deliver close to the required power at 674 nm are commercially available but not space qualified (see sect. 6.2.). Based on experience with the ACES/PHARAO laser system we expect that such systems can be further developed and space qualified with relatively modest industrial investment. Several servo loops are used for locking of the lasers to each other, to the cavity, and to the atoms. Two acousto-optic modulators (AOM) serve for the modulation of LL to interrogate the atomic resonance. The corresponding error signal is sent to the third AOM thereby locking LL to the atoms. A fast phase locked loop (PLL) is used to lock LH to LS and to add an offset frequency δ*f* between the two (to mitigate stray light, see section 3.3.4.2.) and for the frequency modulation of the FSK (or PSK) data transmission (section 3.3.5.). Another PLL serves for the heterodyne detection of the beat between the incoming signal and LL, which provides the science data (frequency difference) and gives access to the data modulated onto the frequency of the incoming signal. The incoming and outgoing channels have linear orthogonal polarizations to minimise interference (stray light), hence the use of the polarizing beam splitter (PBS) separating the two. The RF signals for the AOMs, the PLL loops, and for data time tagging are delivered by an on board quartz USO (*e.g.* the one developed for the ACES project), locked to the hyperfine transition of the atoms used in the accelerometer (see sect. 3.1.) to avoid long term drifts with an uncertainty of <$10^{-12}$ in relative frequency.

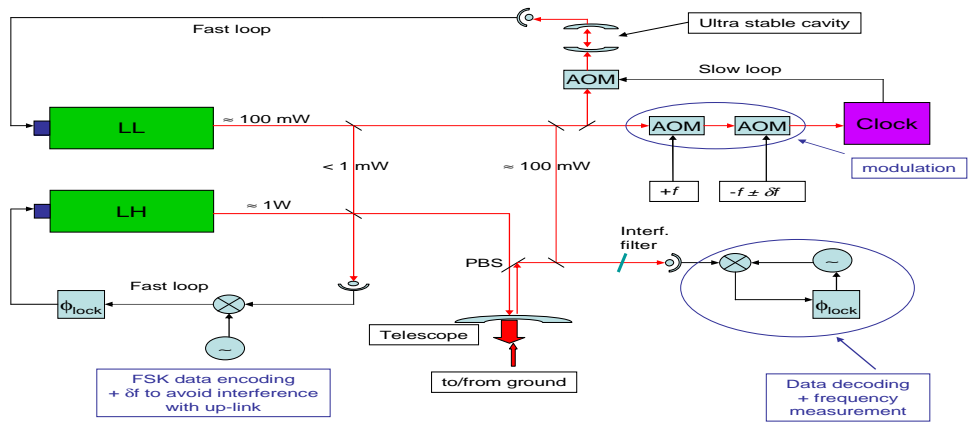

**Fig. 3-1:** Principle of the laser set-up for DOLL

### 3.3.3. Ground segment
The ground segment for DOLL will consist of several (min. 3 to ensure permanent coverage) laser tracking stations. Present satellite and lunar laser ranging stations are well adapted for this purpose, but will require upgrades concerning the optical set-up (wavelength, polarization control, pointing accuracy, adaptive optics) and need to be equipped with high performance optical clocks. Presently several LLR stations are equipped with 1.5 m diameter telescopes (OCA, Matera,…) which are sufficient for DOLL. However, the phase coherent detection requires adaptive optics methods to ensure phase coherence over the complete aperture. Additionally the 3.5 m telescope of the US Apache Point LLR station could provide a valuable high power addition to the ground segment. The minimal ground station laser setup (the one considered in the following sections) would be a symmetric replica of the space set-up, but including a variable control of the orientation of the linear polarization in order to ensure orthogonality between the incoming and outgoing signals on board the S/C and on the ground, and provision to accommodate the large Doppler shift (up to 70 GHz) between the nominal clock wavelength (674 nm) and the emitted/received signal. Upgraded scenarios could use higher emission power (> 10 W, see sect. 6.2.), which in turn would mitigate detection, stray light (on S/C) and pointing issues (on ground). The ground station optical clocks could be either $Sr^+$ ion clocks as for the S/C but also any other optical clock equipped with a fs frequency comb allowing the frequency comparison with the 674 nm nominal wavelength.

### 3.3.4. Error sources
The main error sources affecting the performance of DOLL are related to power and pointing issues, stray light and effects of the Earth's atmosphere. They are discussed in the following sub-sections bearing in mind the overall performance discussed in section 3.3.1.

*3.3.4.1. Power and shot noise:* The power at the centre of the received Gaussian laser beam is given by

$$P_{rec} = \frac{2 A_{rec}}{\pi w(r)^2} P_{emit} \approx \frac{2 A_{rec} A_{emit}}{\lambda^2 r^2} P_{emit} \qquad (3\text{-}1)$$

where $A$ and $P$ are the telescope area and power at emission and reception, $w(r)$ is the beam waist at distance $r$ and $\lambda$ the wavelength. For SAGAS at a distance of 30 AU with 1W emitted power and a 1.5 m ground telescope this corresponds to $P_{rec} = 4.9 \times 10^{-14}$ W = 166000 photons/s. Atmospheric attenuation for a site at 2000 m altitude leads to a loss of typically 30% (between 4% and 40% depending on elevation and weather conditions). Pointing errors contribute exponentially due to the Gaussian profile of the beam. For a $1.5 \times 10^{-6}$ rad (=0.3") S/C pointing error this corresponds to another loss of about 71%. Finally, we allow for a further 35% loss in the instrument, leaving a total power at reception of $6.5 \times 10^{-15}$ W = 22000 photons/s *i.e.* a S/N power of 43 dB (in a 1 Hz band) for the quantum noise limited heterodyne measurement. The corresponding photon shot noise PSD on the frequency measurement is only about $S_y(f) = 1.5 \times 10^{-34} f^2$/Hz, many orders of magnitude below the clock noise ($10^{-28}$/Hz) at the low ($< 10^{-2}$ Hz) frequencies of interest to SAGAS.

*3.3.4.2. Stray light:* Stray light issues can be separated into coherent and incoherent stray light. Incoherent stray light can be overcome by using a narrow band filter and by supplying sufficient power from the local laser to recover a nearly quantum limited coherent photon detection. Coherent light (within the bandwidth of the heterodyne detection PLL) needs to be reduced below the signal power level.

Direct radiation from the Sun at 1 AU has a typical power spectral density of ≈0.07 Wm$^{-2}$cm around the wavelength of interest. Narrow band pass interference filters (*e.g.* the ones used for ACES/PHARAO) show a band-width of about 0.3 nm whilst allowing 90% in-band transmission. At the first occultation (expected at about 2 AU) this leaves about 0.1 Wm$^{-2}$ in-band stray light when pointing directly toward the Sun, and much less for subsequent occultations or when pointing away from the Sun (most of the time). For the S/C this corresponds to about 14 mW power entering the telescope, of which only the ratio "photodiode area" to "Sun spot area" on the focal plane is detected. This is in any case sufficiently small to be overcome by the power from the local oscillator. Another source of incoherent stray light could be the spontaneous emission from the semiconductor amplifier of the high power (1W) laser for emission. Measurements at SYRTE on such systems at the Rb wavelength (780 nm) show a spectral density of such radiation of about $10^{-12}$ W/Hz. Taking into account the 0.3 nm filter, and estimating that $10^{-10}$ to $10^{-7}$ of that radiation is scattered by the optics and the telescope (see below) this leaves < $10^{-8}$ W incident on the detector, which is totally negligible with respect to LL power.

Coherent stray light passing the filter of the heterodyne detection PLL (≈ 1 kHz) is potentially more difficult to deal with, as it needs to be reduced to below the power of the incoming signal (6x10$^{-15}$ W @ 30 AU). Concerning the light from the outgoing beam scattered from the optics (PBS mainly) and the telescope the main reduction of the effect of coherent stray light is achieved by offsetting the high power laser (LH) from the incoming signal frequency (clock frequency) by $\delta f$ = 1 to 10 GHz using a PLL driven by the local RF signal. This is possible because of the $10^{-12}$ accuracy of the local microwave locked to the atoms in the accelerometer, which means that $\delta f$ can be generated with mHz accuracy, sufficient for the $10^{-17}$ uncertainty on the optical frequency measurement. Typically, an ECDL is characterised by a (white) noise spectrum corresponding (in its wings) to a Lorentzian peak of about 10 kHz width. Then the PSD at 1 GHz from the line centre (*i.e.* at the frequency of LL and the incoming laser) is about $10^{-11}$ of the peak PSD, corresponding to $10^{-8}$ W in the 1 kHz PLL filter. The remaining reduction by another $10^{-7}$ needs to be achieved by optimised design of the telescope and optics and by making use of the orthogonal polarizations. A rough study by EADS-SODERN estimates the detected stray light from the outgoing laser in the $10^{-7}$ to $10^{-9}$ range. So we are confident that the required $10^{-15}$ W limit on coherent scattered light from the outgoing laser can be achieved. Sunlight within the 1 kHz filter is further reduced by the ratio of "laser spot size" to "Sun spot size" on the focal plane, which leaves about $10^{-17}$ W. Similarly spontaneous emission from the semiconductor amplifiers contributes $10^{-9}$ inside the 1 kHz PLL bandwidth of which $10^{-7}$ to $10^{-9}$ are scattered by the telescope and optics. Coherent stray light is therefore more than an order of magnitude below the signal power.

*3.3.4.3. Earth atmosphere:* The main limiting effect on the performance of DOLL will be fluctuations due to the Earth's atmosphere. At high frequency (around the 1 kHz of the detection PLL) these can lead to frequent cycle slips and/or loss of lock. At low frequency atmospheric effects lead to phase fluctuations that can affect directly the observables and corresponding science measurements.

High frequency fluctuations are due to atmospheric turbulence and have been measured using optical stellar interferometry [33]. The results are reported in terms of the path delay structure function defined as $D(\tau) \equiv \langle [x(t+\tau)-x(t)]^2 \rangle$ with "$\langle . \rangle$" denoting an ensemble average, and $x$ the atmospheric optical path delay. For $\tau$ between 10 ms and a few seconds the observed $D(\tau)$ follows a power law with a slope around 1.5 . At 10 ms typical values are $D$(10 ms) ≈ 5x10$^{-14}$ m$^2$ [33]. This corresponds to fluctuations with an amplitude of about a third of the 674 nm wavelength of DOLL at 10 ms and to about 6% of the wavelength at 1 ms. Thus cycle slips at the 1 kHz bandwidth of DOLL are unlikely, but cannot be fully excluded, in particular during periods of large turbulence, wind velocities etc… Furthermore, intensity fluctuations, although mitigated by the adaptive optics and appropriate electronics (logarithmic amplifiers) can further increase the likelihood of loosing lock.

At longer integration times fluctuations are dominated by white phase noise due to turbulence, and slowly varying effects due to the evolution of global parameters (temperature, pressure, humidity). The structure function $D(\tau)$ undergoes a transition to white noise ($D(\tau)$ ≈ const.) above a characteristic scale. In the results of [33] this transition occurs between $\tau$ = 10 s and $\tau$ = 100 s with $D(\tau)$ reaching a plateau at about 4x10$^{-10}$ m$^2$. With a data rate of 0.01 Hz for DOLL this corresponds to a variance of 2x10$^{-10}$ m$^2$ of the fluctuations of DOLL data. Additionally, we allow for slow variations, roughly linear over each individual 6 h continuous observation, with an amplitude of 1 mm, corresponding to typical modelling errors of the tropospheric delay from global parameters as used in SLR [34], and a roughly linear evolution of those parameters over 6 h. This corresponds to a white frequency noise with a variance of about 2.4x10$^{-32}$ in relative frequency at 6 h averaging.

To model the corresponding overall atmospheric noise, we generate data (100 s points) with a white phase noise corresponding to 2 10$^{-10}$ m$^2$ delay fluctuations, form the derivative, and then add 6 h constants randomly distributed with 2.4x10$^{-32}$ variance in relative frequency. The resulting PSD in relative frequency corresponds to $S_y(f) \approx 1.8\times10^{-23} f^2$ /Hz for $f > 3\times10^{-4}$ Hz corresponding to the white phase noise from turbulence. It then increases with decreasing frequency reaching a plateau at $S_y(f) \approx 1\times10^{-27}$ /Hz for $f < 3\times10^{-5}$ Hz corresponding to the added white frequency noise.

Finally, one should bear in mind that for the determination of the frequency difference between the S/C and ground clocks any perturbation that is common to the up and down link cancels to a large extent, when the asynchronous data are combined in an optimal way (see [43] for details).

### 3.3.5. Communication and Ranging

The capacity of a coherent optical channel is given by: $C_{coh} = (\log_2 e) \times B \times \ln(1+R/B)$, where $R$ is the rate of detected signal photons and $B$ is the signal bandwidth. For example, with $R = 22000$ s$^{-1}$ and $B = 500$ Hz, one obtains $C_{coh} = 2700$ bps. Data is encoded onto the laser signals of the up and down link using FSK (frequency shift keying), by modulating the laser frequency via the reference microwave signal frequency in the PLL of LH or PSK (phase shift keying) by modulating the phase of the reference microwave signal. Data is decoded by using a tracking oscillator phase locked on the incoming signal. This allows achieving a data transfer rate of several 10$^3$ bps with error bit rates of 10$^{-4}$ or less, which can be further reduced by standard coding schemes. The detection PLL will track the FSK or PSK modulations, thereby achieving the information transfer while simultaneously acquiring the science data (frequency measurement of the optical carrier). When using FSK or PSK the offset frequency is provided by the local microwave and known at << mHz uncertainty, thus this setup allows frequency transfer at 10$^{-17}$ on the optical frequency. However, this requires that cycle slips due to the modulation and noise are $\leq 10^{-3}$/s. For a PLL the average time $\tau$ between cycle slips can be estimated from the variance $\sigma_\Phi^2$ of the phase error of the phase locked loop i.e. $\tau = \pi/(4Bw) \exp(2/\sigma_\Phi^2)$, where $Bw$ is the bandwidth of the phase locked loop. For instance, a 0.2 rd equiprobable binary phase modulation at 500 Hz, 22000 detected photons/s, a loop bandwidth of 1 kHz leads to $\tau \sim 1000$ s. In summary we expect to be able to transfer several 10$^3$ bps without compromising science measurements. More generally, the modulation bandwidth can be modified during the mission (accounting for the changing signal photon rate with distance) for optimal compromise between data transfer capacity and science measurement (cycle slips).

The ranging uncertainty is determined by the capability of timing the FSK or PSK modulations (*i.e.* measuring the arrival time on the local clock of a particular sequence of code). It is a function of the bandwidth of the phase locked loop, and the signal to noise ratio over the time of integration $\tau$, decreasing as $\tau^{-1/2}$. For a 1 kHz bandwidth the timing instability $\sigma_x(\tau) = 6\ \tau^{-1/2}$ μs with $\tau$ in s. In the long term the uncertainty will be limited by systematic effects *e.g.* instrumental delay uncertainties, residual atmospheric effects, etc... which can be controlled to at least the 10 ns level.

The FSK or PSK timing also characterises the instability with which the on-board USO can be synchronized to ground time-scales, and therefore the uncertainty with which on-board measurements can be dated with respect to a ground time-scale. Note that using the up and down link difference for synchronization (two-way configuration) contributions from orbit errors are largely rejected and play no role at the estimated synchronization uncertainties. Long term limits in this case include also the USO instability over dead-time between observations, about 100 ns (allowing for 1 day dead-time with 10$^{-12}$ uncertainty of the USO locked to the Cs atoms in the accelerometer).

In summary, we estimate that using the FSK or PSK coding at $B = 500$ Hz, ranging instabilities are 1.4 $\tau^{-1/2}$ km (with $\tau$ the integration time in s) with systematic effects limiting at the few m level. Similarly, synchronisation of the USO (and therefore timing of on-board data) will be possible with an overall 100 ns uncertainty during optical link down-time, and about 10 ns otherwise.

### 3.3.6. Power, Mass, Volume

The DOLL power and mass are estimated based on the PHARAO clock (diode lasers, AOM, electronics, RF source) and LISA studies (telescope, optical components etc...). Note also that some of the subsystems (LL, ULE cavity, RF source) are already accounted for in the clock or accelerometer budgets, but are included also here for additional margin and to reflect the early stage of the design.

|  | **LH** | **LL** | **ULE cavity** | **AOM+RF+electronics** |
|---|---|---|---|---|
| Power | 25 W | 3 W | 3 W | 20 W |
| Mass | 89 kg | | | |
| Volume | 30 l (not including telescope) | | | |

**Tab. 3-2:** DOLL budgets.

### 3.3.7. Critical issues, requirements, heritage, and technology development

The main DOLL critical issues are the 0.3" pointing requirement and the laser reliability. They are discussed in more detail in sections 4.3. and 6.2. respectively. Much of the DOLL technology is based on heritage from PHARAO (ECDL, AOM, RF synthesis, Quartz,…), COROT (CCD, pointing) and on LISA technology developments (telescope, optics,…). That ensures a relatively high TRL (Technology Readiness Level) for most of the sub-systems, which is however contingent on the reliability and the development of the high power laser source at 674 nm.

Finally, we note, that the development of DOLL as well as the optical clocks for the DOLL ground stations fits well into current technology drives towards the use of optical clocks and optical communication in NASA and ESA interplanetary mission and DSN.

## 4. SPACECRAFT KEY FACTORS

SAGAS has the challenging task to measure the gravitational field in the Solar System between one and 50 AU. The measurement is conducted by a combination of a laser link over interplanetary distances, a cold atom accelerometer and an atomic clock. The preliminary concept satisfying the requirements put forward by its tasks consists of a 3-axis stabilised spacecraft with excellent pointing accuracy. The power demand of the order of 400 W is fed by two Radioisotope Thermoelectric Generators (RTGs). A bi-propellant propulsion module (PM) is attached to SAGAS that serves to increase the Earth escape velocity and/or to conduct required deep space manoeuvres (see sect. 5.).

### 4.1. Design Drivers

The design drivers for SAGAS arise mainly from the needs of the payload and the requirements during a long journey to the outer Solar System.

A general requirement for the payload of SAGAS is high thermal stability, because the precision instruments will deliver best performance in a stable environment. Specific requirements for the cold atom accelerometer and the interplanetary laser link come on top of this:

- The cold atom sensor has the same need as any conventional precision accelerometer to be placed precisely in the centre of mass of the spacecraft with as little self gravity gradient from the spacecraft in its surroundings as possible. Additionally, the envisaged absolute acceleration measurement at $5 \times 10^{-12}$ m/s$^2$ requires that any self gravity bias at the position of the accelerometer be known with that same level of accuracy. The most sensitive axis for the accelerometer science measurements is along the telescope axis, which allows some room for optimisation. As an example, self gravity onboard the MICROSCOPE satellite at the location of the accelerometers is of order $10^{-10}$ - $10^{-9}$ m/s$^2$, with a gradient of about $10^{-11}$ s$^{-2}$. For the projected SAGAS uncertainty this implies knowing the mass and mass distribution of the S/C at the $10^{-2}$ to $10^{-3}$ level. Carefully balanced S/C design could further reduce this requirement.

- The laser link is a two-way asynchronous connection. In order to establish a link over the 50 AU maximum distance to Earth the downlink needs to have an extremely narrow beam that will cover only a small fraction of the Earth during nearly all of the journey. Only at 50 AU the beam diameter will reach the magnitude of the Earth's cross section. The narrow beam divergence implies a pointing requirement of 0.3" (see section 3.3.4.1.). As a further complication the uplink and downlink will have to point into different directions due to the long two-way light time of up to 13.8 h. These requirements will drive the AOCS.

The journey to the outer Solar System puts forward the following requirements to the system:

- A power supply that is still functional at low insolation and that satisfies the considerable power demands of the SAGAS payload of about 200 W.

- A robust communication system suited for the long distance as a backup to the laser communication system.

- A thermal control that can cope with a factor 2500 change in impinging solar power.

In the following sections the solutions for each of the drivers, identified above, are addressed in the context of the relevant subsystem. First the configuration is discussed addressing the accommodation of the cold-atom accelerometer. Then the AOCS system is discussed addressing the issue of the laser-link pointing. Finally the power system, the telecommunication system and the thermal control are outlined.

### 4.2. Configuration

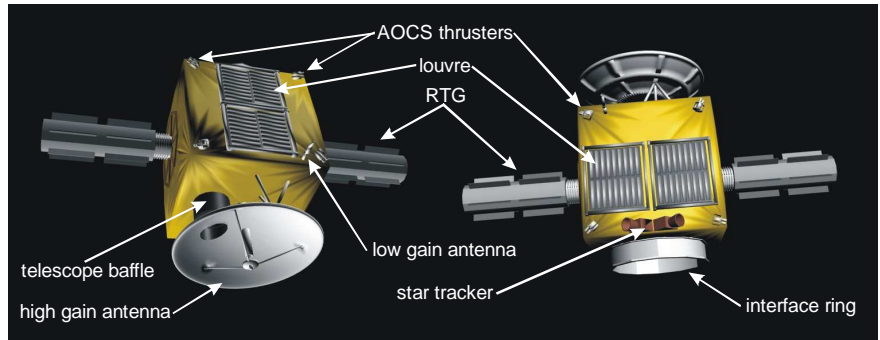

**Fig. 4-1:** Configuration concept of the SAGAS spacecraft. The design shows a variant with 2 GPHS RTGs because the details of the ASRG are not yet available

The SAGAS spacecraft is a three-axis stabilised platform. One side is permanently pointing towards Earth. This side carries both the high gain antenna (HGA) and the Telescope for the laser link. The opposite panel of the S/C carries the adapter ring towards the propulsion module. Two RTG power supplies are placed on the side panels on short struts. The length of the struts is limited by the space in the fairing and by the launch loads. The longest possible struts for the selected launcher should be chosen in order to minimise the view factor of the RTGs towards the S/C side panels. This serves to minimise the backscattering of thermal radiation from the spacecraft hull that contributes an unwanted source of acceleration systematics. For the interior configuration care has to be taken in order to minimize the magnitude of centre of mass movements. For this reason the hydrazine monopropellant for AOCS is distributed over three tanks that are placed towards the sides of the bus and symmetrically with respect to the centre of mass.

### 4.3. AOCS
The SAGAS spacecraft needs to be three-axis stabilized in order to fulfil the pointing needs of the laser link. The attitude control system consists of sensors and actuators. As sensors a Sun sensor is foreseen for initial attitude acquisition in LEOP and safe mode. Star trackers and an inertial measurement unit are used for attitude determination for coarse pointing, e.g. during trajectory manoeuvres. During science mode the attitude determination is supported by a feedback from the detector of the uplink laser signal. This detector is designed as a quadrant photodiode and can hence provide fine pointing information with respect to the uplink direction by differential wavefront sensing. The same feedback from the photodiodes to the AOCS is foreseen for LISA.

Standard reaction wheels of 12 Nms capacity and a set of 12 (+12 redundant) 10 N thrusters are foreseen as actuators. Wheels of a sufficient lifetime for SAGAS are available for instance from Rockwell Collins, Germany. In all nominal modes the attitude control of SAGAS relies on reaction wheels. The fine-pointing at the required level using conventional reaction wheels and a feedback loop from the payload is currently being demonstrated by the COROT spacecraft and is hence also foreseen for SAGAS. It was found that the low frequency vibrations induced by the reaction wheels are compatible with the accelerometer and optical cavity vibration requirements (see sect. 3.1.6.). For safe mode, the attitude control will rely on the thrusters. The thrusters are also used for wheel desaturation and orbit control. The major orbit control manoeuvre to be accomplished will be the targeting before the Jupiter swingby for which 80 m/s of ΔV have been allocated.

A particular challenge for the AOCS, and indeed SAGAS as a whole, is maintaining lock to the incoming laser, and initial acquisition of that lock. The solution to this problem lies in the combined use of the CCD (FoV of 1') that the S/C telescope is equipped with and the quadrant diode (0.3" FoV) used for the heterodyne detection (a combination of the COROT and LISA methods for fine pointing). Initial acquisition is achieved in the following way:
1. The S/C establishes the right pointing towards a ground station G with respect to the Earth's image in the CCD using onboard information about the station position at a given time with respect to the Earth's limb and starts emitting, applying the necessary point ahead angle so that the signal arrives at G (or another) ground station (see below for requirements).
2. The Gs all stare with their CCDs (large FoV) waiting for an incoming signal. On reception they lock onto it, apply the necessary point ahead angle and emit back.
3. The S/C thus receives a signal back, already in its quadrant diode FoV, and locks onto it.

Note, that lock remains established even when Gs change due to Earth's rotation because the S/C "knows" where the different Gs are, and thus switches from one to the next as required (once lock is established the S/C disposes of the incoming laser additionally to the CCD image when switching Gs). It also switches the point ahead angle (not synchronously with the reception switch, but offset by the light travel time) so Gs always have an arriving laser to lock to (they do not need to scan for the S/C with their CCD). The procedure requires that the CCD FoV and the quadrant diode FoV be aligned to better than 0.3". LISA studies have shown that alignment to below 0.2" is possible. For SAGAS, additionally, we have the possibility of post launch calibration of that alignment using the large CCD image of the Earth and large incoming laser power in the early mission phase. Secondly, the CCD needs to be able to resolve a target on the Earth image corresponding to the 0.3" diode FoV, *i.e.* about 200 km @ 1 AU and 6500 km @ 30 AU, which should be possible given COROT performance.

### 4.4. Power Subsystem

For a mission to several tens of AU a radio thermal power supply is mandatory. The payload and system power demands are 400 W in total, both in cruise mode with PM and in science mode. For the preliminary design, four ASRG RTGs which are currently being developed by Lockheed-Martin in the US are considered. The development plan by NASA and the US Department of Energy for RTGs foresees that these Stirling Radioisotope Generators will be available in 2009 and hence well in time for the Cosmic Vision timeframe. The performance assumptions for the ASRG are based on [35]. The specific type of RTGs is not important for SAGAS. For instance, a set of two current US GPHS RTGs would also be a suitable choice that fulfils the power and lifetime requirements of SAGAS.

In fact, the ASRG Stirling generators are unfavourable for the SAGAS application because their moving pistons will cause vibrations. These are, however, not considered critical because the ASRG have two counter-moving pistons to reduce vibrations and will be equipped with an adaptive vibration reduction system [36]. Due to the fixed frequency of the piston motion the vibrations caused by them can furthermore easily be damped and identified in the accelerometer data. Hence the use of the ASRG is considered unproblematic, although the use of conventional RTGs with thermocouples would be preferable, if they are available for the timeframe of the SAGAS implementation.

### 4.5. Radio-Frequency Telecommunication System

An X-band communication system is foreseen for telemetry tracking and command. The radio link is mandatory because during critical mission phases such as orbit manoeuvre and in safe mode the laser link will not be available. For LEOP two low-gain antennas with hemispherical coverage are foreseen. For deep space communications a high-gain antenna (HGA) is baselined. It is assumed that the communication can be established via the HGA also in safe mode relying only on Sun-sensor information using a conscan manoeuvre. Hence currently a medium gain antenna is not foreseen. Since the current design is not mass critical this decision may be revisited. The accommodation of the HGA is constrained by the accommodation of the 40 cm aperture telescope. The options were identified:

- A rather small HGA of only 1.1 m diameter could be used to accommodate the HGA side by side with the telescope. Combined with a transponder with 27.5 W transmit power such as it will be used in BepiColombo a data rate of 50 bps is achievable over 50 AU which is sufficient taking into account that the channel for the science data and regular telemetry is via the laser link.
- A larger HGA could be used if a 45 cm cut-out for the telescope is allowed in the dish. Since the loss in antenna gain will be roughly proportional to the area of the cut-out this solution is viable if the antenna is sufficiently large. Assuming the same transponder performance as above and 2.2 m diameter antenna as it was for instance used for Rosetta a data rate of 180 bps is achievable at the end of mission. In order to limit the impact on the antenna pattern a cut-out located close to the centre of the HGA is preferable. Due to the small field of view of the telescope the placement of the antenna feed does not pose a particular problem.

The larger antenna with the cut-out is the currently preferred solution due to its higher performance. A detailed assessment of the effect of the cut-out on the antenna pattern is however needed to finalize the decision.

## 4.6. Operations Concept

The two most frequently used spacecraft modes of SAGAS will be the Science Mode and the Normal Mode. Both modes will be reflected in the AOCS modes, power modes, and operational schemes. Normal Mode will be used for orbit control and when science operations are interrupted. Science mode will be used when the cold-atom accelerometer and/or the laser link are operating.

The Normal Mode will be very similar to that used during cruise for a typical interplanetary mission such as MarsExpress. During Normal Mode, the Mission Operations Centre will have exclusive control of AOCS and operations scheduling. Communications will be via the radio-frequency link. This mode will in particular be used during orbit manoeuvres and for their preparation. The AOCS will be coarse pointing via the star trackers and inertial measurement units.

The Science Mode will be used during most of the mission, in particular when the spacecraft is coasting in deep space. In Science Mode the laser link will be established and spacecraft AOCS will be fine-pointing using the laser link information. In this mode the mission operations and science operations will be deeply intertwined and it will be necessary that several typical mission operations tasks are under direct supervision of the Science Operations Centre. In particular, this needs to be the case for attitude control, acquisition of the laser beam and scheduling of those spacecraft operations that could interfere with the science tasks through thermal or mechanical disturbances. In Science Mode the primary TT&C connection will be via the laser link.

In order to assure spacecraft safety, the spacecraft AOCS shall be largely autonomous in Science Mode. In addition, a reliable FDIR is required and a robust Safe Mode that will be triggered by watchdogs of the on-board autonomy system. For minimal interruptions of science operations, a largely autonomous recovery from Safe Mode is desirable. It still needs to be assessed if wheel desaturation shall be implemented also as a submode of the Science Mode, in which case it would have to be carried out autonomously by the spacecraft or only for Normal Mode, for which both, autonomous and non-autonomous wheel desaturation, would be possible.

## 4.7. Thermal Control

The SAGAS thermal control is designed to maintain constant ambient temperature for the payload over the whole mission duration, while the impinging power ranges between 950 W at the Earth flyby and a 0.4 W at end of mission. This task can still be accomplished in a straightforward manner by employing louvers. The hot case of the Earth swing-by together with the attainable louver area of ~2.7 m² set the nominal spacecraft temperature to 35°C. This temperature is maintained throughout the mission by the power dissipated by the spacecraft and the changing opening factor of the louvers. Gilded Kapton is foreseen to cover the outer surface of the spacecraft in order to achieve a low emissivity and absorptance. Compensation heaters are mostly unnecessary. Only for the payload compensation heaters at the level of 1/3 of the ON power are foreseen to maintain thermally stable conditions also if the instruments are in standby mode. As an alternative to louvers switchable internal and external power dumpers as on Ulysses could be used.

## 4.8. Propulsion System

The propulsion system of SAGAS serves to carry out trajectory manoeuvres, wheel desaturation and attitude control in safe mode. The trajectory design for SAGAS foresees a trajectory which requires a large deep space manoeuvre of the order of 600 m/s. The large propellant demand for this manoeuvre is conflicting with the requirement to have a well known centre of mass position. The requirements are reconciled by carrying out the large deep space manoeuvre by a propulsion module that is afterwards jettisoned. Only after the jettisoning of the propulsion module the accelerometer takes up its regular duty. A suitable candidate for the propulsion module could be identified in the EUROSTAR series of geostationary platforms, a well known offspring from which is the LISAPathfinder propulsion module. These bi-propellant propulsion modules reach a specific impulse of up to 325 s and are hence well suited to minimise the propellant penalty for the deep space manoeuvre. No lifetime issues exist for this heritage because the EUROSTAR platform is qualified for 15 years of in-orbit lifetime. This series of platforms is under continuous development and is hence likely to be still available also for SAGAS. The key parameters of the propulsion module are given in Table 4-4, below. They were derived from the critical design review data of the LISA Pathfinder PM by a conservative scaling taking into account the different mass of the spacecraft and the propellant.

The only major remaining manoeuvre to be carried out by the SAGAS spacecraft itself is the targeting for the Jupiter swingby which will be of the order of 80 m/s. For this purpose and for attitude control a set of 8 (+8 redundant) 10 N thrusters is foreseen.

**Tab. 4-4:** Propulsion module key parameters based on scaling of the LISA Pathfinder PM

| PM wet mass [kg] | PM dry mass [kg] | Propellant mass [kg] | Maximal ΔV [m/s] | ΔV margin | Isp [s] |
|---|---|---|---|---|---|
| 1134 | 174 | 970 | 1865 | 5% | 325 |

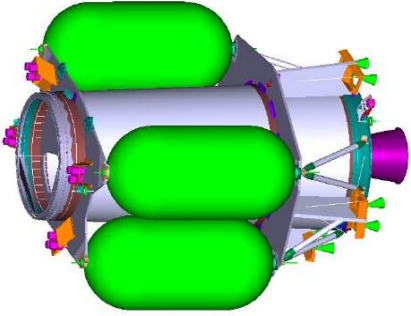

**Fig. 4-2 :** Potential propulsion module for SAGAS based on the EUROSTAR/LISA Pathfinder Propulsion Module heritage

# 5. MISSION PROFILE

SAGAS will deliver precision data on the gravitational field in the Solar System from 1 AU to 50 AU. The total mission duration is 15 years nominal and 20 years extended mission. The spacecraft wet mass under consideration will be 1000 kg. The trajectory foresees a Jupiter gravity assist to reach hyperbolic escape velocity. Depending on the chosen launcher, one or more gravity assists in the inner Solar System may be required to reach Jupiter. The objective of the trajectory is to reach the end of mission target, a heliocentric distance of 50 AU in as short time as possible.

## 5.1. Escape Strategy and Launcher Selection

Combined with the sizeable spacecraft mass, a large launcher is the only realistic option. The only currently available European launcher that falls into this category is the Ariane 5 ECA. For a 671 kg spacecraft, it offers a hyperbolic excess velocity of 7 km/s [37]. Unfortunately, this performance is too low to put a spacecraft of the desired mass into a trajectory towards Jupiter. Hence even with an Ariane 5 ECA, a gravity assist in the inner Solar System will be required to put the spacecraft on the desired trajectory.

The situation could considerably improve with the advent of the Ariane 5 ECB, which will feature the re-ignitable Vinci upper stage engine. With the Ariane 5 ECB, a 1000 kg spacecraft could reach an escape velocity of 9.5 km/s, which would be more than sufficient for a direct transfer to Jupiter. It is likely that the Ariane 5 ECB will be available for the Cosmic Vision time frame.

Other currently available launcher options that would allow a direct transfer to Jupiter are listed in Tab. 5-1. These launchers could be considered if the mission is realised in collaboration with other space agencies.

| Launcher options | Hyperbolic excess velocity [km/s] for 1000 kg payload |
|---|---|
| Ariane 5 ECA | ~ 6.5 km/s |
| Ariane 5 ECB | ~ 9.5 km/s |
| Atlas V / Star 48V | ~ 10.7 km/s |
| Delta IV Heavy / Star 48B | ~ 10.0 km/s |
| Proton M / Breeze M | ~ 8.1 km/s |

**Table 5-1: Launcher options for SAGAS**

For the Ariane 5 launcher the use of a propulsion module (PM) improves the escape performance. Tab. 5-2 gives the performance for the two Ariane 5 variants using the US Star 48B propulsion module and bi-propellant PM. The estimates for the bi-propellant PM are based on the masses and performances of the Eurostar platform and the LISAPathfinder PM that has been derived from it. The design of the PM has been optimised for achieving maximal hyperbolic excess velocity for a 1000 kg payload (The PM is further discussed in Sect. 4.8.). For the combination of Ariane 5 ECA and a Star 48B kick stage, no improvement in performance is achieved.

| Launcher options | Hyperbolic excess velocity [km/s] for 1000 kg payload |
|---|---|
| Ariane 5 ECA / Star 48B | ~ 6.2 km/s |
| Ariane 5 ECB / Star 48B | ~ 10.5 km/s |
| Ariane 5 ECA / Eurostar | ~ 6.7 km/s |
| Ariane 5 ECB / Eurostar | ~ 10.5 km/s |

Tab. 5-2: Propulsion module options

From this survey, it is clear that the Ariane 5 ECB, the American Atlas V and Delta IV and the Russian Proton allow a direct transfer to Jupiter if a kick-stage or propulsion module is used. For the Ariane 5 ECA, a direct transfer to Jupiter is not obtainable even with a propulsion module.

**5.2. Trajectory Design**

Amongst the launchers discussed above, only the Ariane 5 ECA is considered in the launcher list of the Cosmic Vision frame [38]. Hence, we consider the Ariane 5 ECA as our preferred launcher option. Unfortunately, this launcher has the lowest performance, and hence requires a more complicated trajectory design to reach the desired heliocentric distance of 50 AU. The following principle options exist to put the spacecraft onto a trajectory towards 50 AU after launch with an Ariane 5 ECA:

- An Earth-Venus-Earth-Earth-Jupiter (EVEEJ) trajectory with four gravity assists. Such trajectories have been employed for the Galileo and Cassini missions
- An Earth-Earth-Jupiter trajectory with two gravity assists. In this option the spacecraft is first put into a resonant orbit with Earth of two or 1.5 years period and a sizeable deep space manoeuvre of ~ 0.5 km/s is conducted at aphelion to amplify the effect of the gravity assist at Earth. Such trajectories are commonly denoted as ΔV-EGA trajectories.

In both options, hyperbolic escape velocity is reached after Jupiter and the SAGAS spacecraft reaches 50 AU in a hyperbolic coast. A final swingby at Saturn after that at Jupiter could considerably enhance the escape velocity. Unfortunately, no good options exist for a Jupiter-Saturn trajectory after 2016 throughout the Cosmic Vision timeframe. Hence this option is not further considered.

In general the ΔV requirement for the EVEEJ trajectory is ~0.7 km/s smaller than that for the ΔVEGA option. Nevertheless for SAGAS the ΔVEGA option is preferred because it avoids the thermally challenging environment of Venus at 0.7 AU heliocentric distance.

For preliminary mission analysis, a global optimisation has been carried out to determine the preferred transfer opportunity with a launch date between 2017 and 2025. The objective was to reach 50 AU in minimal travel time under the combined constraints of the Ariane 5 ECA escape capability and the maximal ΔV capability of the propulsion module for the respective escape conditions. The optimal trajectory with a departure date in the Cosmic Vision timeframe has a launch in March 2019 and leads to a transfer to 50 AU of only 18.83 years.

Not all ΔV provided by the propulsion module is needed for the deep-space manoeuvre during the EGA loop. The remaining ΔV is applied after the Earth flyby in order to increase the departure velocity. A higher departure velocity after the flyby could be reached if the ΔV would be applied during the gravity assist. However the burn time for the manoeuvre of ~850 m/s magnitude will be longer than 1/2 hour even for an uninterrupted burn. It requires a detailed analysis to determine how much of the available ΔV can be applied within the Earth's sphere of influence taking into account operational constraints. Here we take the conservative approach of not considering a powered swingby. A moderate reduction of the available deep space ΔV, e.g. due to an increased mass of SAGAS, would not result in an infeasible trajectory but just in a slightly extended mission duration. The key parameters of the resulting optimal trajectory are given in Tab. 5-3. The trajectory until Jupiter is depicted in Fig. 5-1.

Generally, launch opportunities towards Jupiter open up once a year. The analysis showed that the typical travel times for these opportunities are between 19 and 21 years. Hence, allowing a 10% longer travel time, the launch of SAGAS can take place in any desired year. This flexibility in launch date is a particular benefit of the ΔV-EGA for which only an optimal constellation of Earth and Jupiter is required which arises once a year.

For the other launchers in Tab. 5-1 a direct transfer to Jupiter is possible. This would allow to omit the EGA loop and hence to shorten the mission duration by approximately 2 years.

| Event | Date | Time since launch [days] | Time since launch [years] | Interval between events [days] | Interval between events [years] |
|---|---|---|---|---|---|
| Launch | 22 March 2019 | 0 | 0 | 0 | 0 |
| 1st deep-space manoeuvre | 7 May 2020 | 412 | 1.13 | 412 | 1.13 |
| Earth gravity assist | 15 May 2021 | 785 | 2.15 | 373 | 1.02 |
| 2nd deep-space manoeuvre | 15 May 2021 | 785 | 2.15 | 0 | 0.00 |
| Jupiter Gravity assist | 29 Oct. 2022 | 1317 | 3.61 | 532 | 1.46 |
| Arrival at 50 AU | 21 Jan. 2038 | 6878 | 18.83 | 6346 | 17.37 |

**Tab. 5-3a:** Dates of trajectory milestones of the optimal trajectory

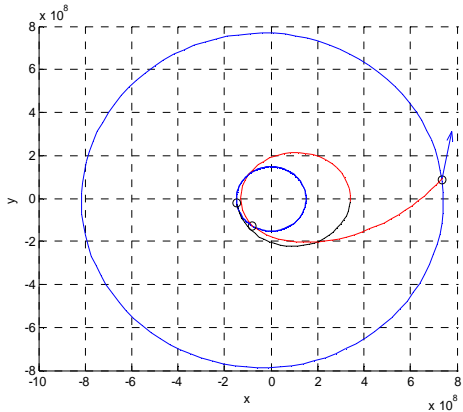

| Parameter | Value |
|---|---|
| Earth departure velocity [km/s] | 5.27 |
| 1st Deep Space ΔV [km/s] | 0.656 |
| 2nd Deep Space ΔV [km/s] | 0.848 |
| Total deep space ΔV [km/s] | 1.504 |
| Velocity after Jupiter swingby [km/s] | 22.64 |

**Tab. 5-3b:** Velocity parameters of the optimal trajectory

**Fig. 5-1:** Optimal ΔV-EGA trajectory for SAGAS through the inner Solar System (red/black). Orbits of Earth and Jupiter in blue. Dimensions in km.

In summary, with the chosen trajectory SAGAS will reach a heliocentric distance of 38.9 AU over the nominal mission duration (15 years) and 53.3 AU over extended mission (20 years).

### 5.3. Ground Segment
For the X-band link the mission will use ESA 15 m ground stations for LEOP operations and ESA 35 m ground stations for deep space communications. During most of the mission, ground contact will be infrequent – less than once a fortnight – because the science and telecommand data are transmitted via the laser link (kbps capacity, see sect. 3.3.). For periods around the gravity assists and the deep space manoeuvres, permanent ground coverage via X-band and laser is desirable. In general no large baseline tracking operations will be required because orbit reconstruction accuracy from the laser ranging will considerably exceed ΔVLBI performances.

The laser link will require a minimum of three dedicated ground stations equipped with 1 m or larger telescopes (see sect. 3.3.), optical clocks and corresponding laser systems (see also sect. 6.2.). The present baseline for such ground stations is to take advantage of the existing structure of Satellite and Lunar Laser Ranging stations, several of which are already equipped with 1.5 m telescopes (OCA, Matera, …). The corresponding stations will require upgrades to make them compatible with SAGAS requirements; in particular, they will require adaptive optics, optical clocks and corresponding laser systems. An alternative option would be to develop dedicated laser DSN stations for SAGAS but also for other deep space missions that will require precise timing, navigation and broadband communication. Both options fit well into the general technological development of laser communication and optical clocks for DSN, carried out presently under ESA and NASA contracts.

### 5.4. Special Requirements
The launcher selection will not only be determined by the launcher's capabilities, but also by the regulations concerning the launch of spacecraft equipped with RTGs. This is currently possible with the Russian and US launchers. However, it is likely that the necessary clearance will be achieved for Ariane 5 in the framework of ESA's Aurora programme (this issue is addressed in more detail in sect. 6.1.).

# 6. KEY TECHNOLOGY AREAS

## 6.1. Platform

The SAGAS spacecraft faces considerable challenges in terms of instrument accommodation, pointing requirements and the need to provide sufficient power in the outer Solar System. The preliminary concept (section 4.) has addressed all theses challenges and has identified feasible solutions. The biggest challenge is the differential wave-front sensing for the locking of the laser beam. The corresponding technology has already been demonstrated for LISA in a laboratory setup. Nonetheless, while considerable experience can be drawn from COROT, GAIA and the technology development for LISA, a detailed analysis of the beam recovery and locking scheme for SAGAS is likely to reveal important differences. Hence a TDA on the attitude for SAGAS during science mode should be initiated timely.

The location requirement of the cold-atom accelerometer in close distance from the centre of mass is comparable to those of GOCE or LISA Pathfinder and hence ample experience on this task will exist in Europe for the implementation of SAGAS.

Otherwise, no particular technology development needs for the platform could be identified. The complete platform of SAGAS uses well developed technology and hence the challenges of SAGAS lie mainly in its payload and operations and not in the platform. Notable exceptions are the Stirling radioisotope generators, which are still under development. These are, however, not considered critical because the type of RTGs to be used is not critical for SAGAS and it can be assumed that an RTG of the power-to-mass ratio of the current GPHS RTGs will, in any case, be available in the US during the Cosmic Vision timeframe.

It is worth stressing that sufficient expertise for the power system design using RTGs and their integration exists in Europe: GPHS RTGs have already been used on the joint ESA/NASA mission Ulysses. For Ulysses, both the power system was designed and the RTGs were integrated by a European company. While the knowledge for the handling of RTG is available in Europe, no corresponding handling and safety regulations for Centre Spatial Guyanais have been established, yet. First steps towards such regulations are currently undertaken for radioisotope heating units within ESA's Aurora programme. However, it is likely that the larger amount of nuclear material present in the RTGs will require additional regulations. Hence, the qualification of CSG for the launch of RTGs should be initiated well ahead of the SAGAS implementation phase. A possible way to circumvent this qualification would be the launch on a US launcher. This option was found attractive from the point of view of trajectory design as well (cf. sect. 5.1.).

## 6.2. Laser Sources for Clock and Optical Link at 674 nm

At 674 nm with the required 1W output, diode laser technology requires the use of an ECDL followed by a tapered amplifier (TA). At present, the maximum output power available at 674 nm using ECDL-TA configuration is 250-300mW, but power upgrades are expected from the development of longer TPA chips. Output power exceeding 1 W is realistic within three years from now. Accelerated ageing tests have been performed with AR-coated 670 nm wavelength ECDLs (90 °C operation temperature), with typical output powers of 10 to 20 mW. An extrapolated value >$10^4$ hours was found in standard operating conditions. No significant dependence of the lifetime on the optical power was found, so that this result can be representative also in case of higher optical power. In standard operating conditions (25 °C) lifetimes exceeding $2.5 \times 10^4$ hours (about 3 years) have been already reported by different customers at 671 nm wavelength. Lifetimes approaching $10^5$ hours are not unrealistic in the future (in fact such lifetime values have been already demonstrated in case of infrared laser diodes for telecom applications) but would require significant developments for the chips and 3-5 years from now. It is thus realistic to expect to be able to cover the complete mission duration with moderate (two or three-fold) redundancy.

An alternative approach based on laser diode technology that could become realistic in the near future is a directly modulated DFB laser (100 mW oscillator) or a DFB laser as master laser within a master oscillator power amplifier system (MOPA) in the case of larger optical power (up to 1 W). DFB lasers offer a significant advantage in comparison with external cavity diode lasers in terms of dimensions and mode-hop free tuning range without any moving mechanical part. The lack of mechanical parts qualifies DFB lasers especially for space applications. Currently, this concept can be realized within the wavelength interval between 730 nm and 1060 nm, but further developments are expected in the next years to extend the emission wavelength towards the visible region.

Although operation of the optical link with 1W ground lasers is certainly possible (symmetric to the down link), it will likely be of advantage to use higher power (> 10 W) on the ground, if available. To that aim a realistic solution was identified using a coherent solid-state device at 674 nm based on intracavity frequency doubling of the $^4F_{3/2} \rightarrow {^4I_{13/2}}$ laser transition at 1348 nm of $Nd^{3+}$ doped crystals. The proposed laser source is based on a slave laser with a thin disk pumping scheme, using a high Nd-doping concentration in a $La_2SiO_5$ crystal (in this case the La ionic radius closely matches the Nd one), injected by a tunable, single-frequency, narrow-linewidth oscillator (500 mW, linear cavity). Using this optical configuration we expect to obtain 30 W at 1348 nm using 100 W pump power at 808 nm and 12-15 W output power at 674 nm by intracavity frequency doubling.

In summary, availability, space qualification, and lifetime of laser sources for the clock and the link (674 nm) is a key technological issue for SAGAS, with a presently low technology readiness. However, given the large technology basis in the field of semiconductor and solid state lasers a rapid development of appropriate laser sources seems likely but should be initiated early.

## 7. CONCLUSION

We have presented a detailed description of the scientific and technological aspects of the SAGAS project, based on the original mission proposal submitted to ESA in June 2007 in response to the Cosmic Vision 2015-2025 call for proposals. The outcome of the ESA selection procedure is now known, and unfortunately SAGAS has not been deemed a priority. However, independently of that outcome, we believe that it is worth pursuing the investigations initiated by the Cosmic Vision call, in order to further study the scientific and technological implications of this type of mission. The present document is a first step towards that goal, allowing public access to most of the presently existing material on SAGAS, thereby providing a basis for further developments.

On the science side, deep space gravity probes are unique opportunities to address some of the most fundamental questions of contemporary physics, related to unification of the fundamental interactions in nature, the nature of gravitation, dark energy and dark matter. By extending experimental tests of gravity to the largest scales attainable by human-made artifacts (size of the Solar System), missions like SAGAS are starting to bridge the gap between observational evidence at relatively short scales ($\approx$ Earth-Moon distance) where all observations confirm present theories, and astronomical (Galaxies) and cosmological scales where agreement between theory and observation comes at the expense of postulating large amounts of dark matter and energy.

Concerning the exploration of the outer Solar System (Kuiper belt, giant planets), SAGAS opens a new and complementary window on such exploration, no longer based on electromagnetic imaging, but on the measurement of the gravitational signatures of the objects to be studied or discovered. The determination of the Kuiper belt mass distribution and total mass are good examples where gravitational measurements are complementary to, and better adapted than "classical" techniques.

The SAGAS payload will include an optical atomic clock optimised for long term performance, an absolute accelerometer based on atom interferometry and a laser link for ranging, frequency comparison and communication. The complementary instruments will allow highly sensitive measurements of all aspects of gravitation via the different effects of gravity on clocks, light, and the free fall of test bodies, thus effectively providing a detailed gravitational map of the outer Solar System, whilst testing all aspects of gravitation theory to unprecedented levels. Table 7-1 provides a summary of the SAGAS science objectives.

| Science Objective | Expected Result | Comments |
|---|---|---|
| Test of Universal Redshift | $1 \times 10^{-9}$ of GR prediction | $10^5$ gain on present |
| Null Redshift Test | $1 \times 10^{-9}$ of GR prediction | $10^3$ gain |
| Test of Lorentz Invariance | $3 \times 10^{-9}$ to $5 \times 10^{-11}$ (IS or "time dilation" test) | $10^2$ to $10^4$ gain fct. of trajectory |
| PPN test | $\delta(\gamma) \leq 2 \times 10^{-7}$ | $10^2$ gain may be improved by orbit modelling |
| Large Scale Gravity | - Fill exp. data gap for scale dependent modif. of GR<br>- Identify and measure PA to < 1% per year of data | Different observation types and large range of distances will allow detailed "map" of large scale gravity |
| Kuiper Belt (KB) Total Mass | $\delta M_{KB} \leq 0.03\ M_E$ | Dep. on mass distribution and correlation of clock meas. |
| KB Mass Distribution | Discriminate between different common candidates | Will contribute significantly to solution of the "KB mass deficit" problem |
| Individual KB Objects (KBOs) | Measure $M_{KBO}$ at $\approx 10\%$ | Depending on distance of closest approach |
| Planetary Gravity | -Jupiter Gravity at $\leq 10^{-10}$<br>-Study Jupiter and its moons | $10^2$ gain on present for Jupiter idem for other planet in case of 2nd fly-by |
| Variation of Fund. Const. | $\delta\alpha/\alpha \leq (2 \times 10^{-9})\ \delta(GM/rc^2)$ | 250-fold gain on present |
| Upper limit on Grav. Waves | $h \leq 10^{-18}$ @ $10^{-5}$ to $10^{-3}$ Hz | Integration over one year |
| Technology Developement | Develops S/C and ground segment technologies for wide use in future missions (interplanetary timing, navigation, broadband communication,…) | |

**Tab. 7-1:** Science objectives of SAGAS (see sect. 2. for details). Red = Fundamental physics, Blue = Solar System science. (GR: General Relativity, PA: Pioneer Anomaly, IS: Ives-Stilwell, $M_E$: Earth mass).

The quantitative estimates in Tab. 7-1 are obtained using rough estimates, as described in the different sub-sections of section 2. Further studies are required to underpin the estimated noise sources and uncertainties given in sections 2 and 3. A particular point of interest is also the possibility to combine the on-board and ground measurements in different ways (sum or difference, variable time delay between up and down link), which can be tailored and optimized for each particular science objective. As a consequence the data analysis can be optimized as a function of the spectral signature of the signal searched for, with potentially increased reliability. Studies on such data analysis strategies are in progress and will be the subject of a forthcoming publication.

In summary, SAGAS opens the way towards the experimental investigation of some of the most puzzling questions of contemporary physics and towards a new window for the exploration of the outer Solar System. These "phenomena of the very large" are explored using the "technology of the very small" (quantum sensors), illustrating the discovery potential of the combination of the two domains.

**ANNEX: List of Acronyms**

S/C: Spacecraft
PPN: Parameterized Post Newtonian
PSD: Power Spectral Density
CMB: Cosmic Microwave Background
DSN: Deep Space Network
USO: Ultra Stable Oscillator
TRL: Technology Readiness Level ( http://sci.esa.int/science-e/www/object/index.cfm?fobjectid=37710 )
SNR: Signal To Noise Ratio
SLR: Satellite Laser Ranging

LLR: Lunar Laser Ranging  
CCD: Charge Coupled Device  
FoV: Field of View  
RTG: Radioisotope Thermoelectric Generator  
ECDL: Extended Cavity Diode Laser  
DFB: Distributed Feedback  
MOPA: Master Oscillator Power Amplifier  
PM: Propulsion Module  
HGA: High Gain Antenna  
LEOP: Launch and Early Orbit Phase  
AOCS: Attitude and Orbit Control System  
GPHS: General Purpose Heat Source  
ASRG: Advanced Stirling Radioisotope Generator